\documentclass[doublespace]{article}
\usepackage{arxiv}

\usepackage[utf8]{inputenc} 
\usepackage[T1]{fontenc}    
\usepackage{hyperref}       
\usepackage{url}            
\usepackage{booktabs}       
\usepackage{amsfonts}       
\usepackage{nicefrac}       
\usepackage{microtype}      
\usepackage{lipsum}		
\usepackage{graphicx}

 \usepackage{moreverb}
\usepackage{pdflscape}
\usepackage{graphicx}
\usepackage{enumitem}

\usepackage{array}
\usepackage{mathtools} 
\usepackage{chngpage}

\DeclareMathAlphabet{\mathbfsf}{\encodingdefault}{\sfdefault}{bx}{sl}

\usepackage{tikz}
\usetikzlibrary{mindmap}			
\usetikzlibrary{arrows,arrows.meta, calc,positioning, fit}
\usetikzlibrary{shapes,decorations, decorations.pathreplacing}
\definecolor{mylightskyblue}{rgb}{0.53,0.81,0.98}
\definecolor{mydodgerblue3}{rgb}{.09,0.45,0.80}
\definecolor{mynavy}{rgb}{0, 0, 0.5}

\usepackage{tabularx}

\usepackage{amsmath, stackrel, bm}
\usepackage[font=normalsize]{subcaption}
\usepackage{multirow}
\usepackage{acro}
\acsetup{first-style=long-short}
\DeclareAcronym{ATE}{
  short = ATE,
  long  = average treatment effect
}
\DeclareAcronym{ATT}{
  short = ATT,
  long  = average treatment effect of the treated 
}
\DeclareAcronym{CART}{
  short = CART,
  long  = Classification And Regression Tree
}
\DeclareAcronym{DAG}{
  short = DAG,
  long  = directed acyclic graph
}
\DeclareAcronym{IPTW}{
  short = IPTW,
  long  = Inverse Probability of Treatment Weighting
}
\DeclareAcronym{GBSG}{
  short = GBSG,
  long  = German Breast Cancer Study Group
}
\DeclareAcronym{MFP}{
  short = MFP,
  long  = Multivariable Fractional Polynomial
}\DeclareAcronym{MOB}{
  short = MOB,
  long  = Model-based Recursive Partitioning for Subgroup Analysis
}
\DeclareAcronym{MSE}{
  short = MSE,
  long  = Mean Squared Error
}
\DeclareAcronym{palmtree}{
  short = palmtree,
  long  = partially additive (generalized) linear model tree
}
\DeclareAcronym{RFS}{
  short = RFS,
  long  = Relapse-free survival
}\DeclareAcronym{PI}{
  short = PI,
  long  = permutation importance
}

\usepackage[numbers, super]{natbib}
\newcommand\BibTeX{{\rmfamily B\kern-.05em \textsc{i\kern-.025em b}\kern-.08em
T\kern-.1667em\lower.7ex\hbox{E}\kern-.125emX}}

\title{Statistical Plasmode Simulations - Potentials, Challenges and Recommendations}

\author{{Nicholas Schreck\thanks{Both authors contributed equally.}} \\
	Biostatistics Division\\
	German Cancer Research Center\\
    D-69120 Heidelberg, GERMANY \\
	\texttt{nicholas.schreck@dkfz-heidelberg.de}
	\And
	{Alla Slynko\footnotemark[1]} \\
	Department of Statistics and Actuarial Science\\
	University of Waterloo\\
	Ontario, N2L 3G1, CANADA \\
    \And
	{Maral Saadati} \\
	Freelance Statistician\\
	Saadati Solutions\\
	D-68526 Ladenburg, GERMANY \\
	\And
	{Axel Benner} \\
	Biostatistics Division\\
	German Cancer Research Center\\
    D-69120 Heidelberg, GERMANY \\
}
\newcommand{\headeright}{N. Schreck et.~al.}

\renewcommand{\shorttitle}{Statistical Plasmode Simulations}

\begin{document}
\maketitle







\begin{abstract}
Statistical data simulation is essential in the development of statistical models 
and methods as well as in their performance evaluation. 
To capture complex data structures, in particular for high-dimensional data, 
a variety of simulation approaches have been introduced including parametric 
and the so-called plasmode simulations. 
While there are concerns about the realism of 
parametrically simulated data, it is widely claimed that plasmodes come very close to 
reality with some aspects of the ``truth'' known.  
However, there are no explicit guidelines or state-of-the-art on how to 
perform plasmode data simulations. 
In the present paper, we first review existing literature and 
introduce the concept of statistical plasmode simulation.
We then discuss advantages and challenges of statistical plasmodes and provide
a step-wise procedure for their generation, including key steps to their implementation 
and reporting. 
Finally, we illustrate the concept of statistical plasmodes as well as the proposed 
plasmode generation procedure by means of a public real RNA dataset on breast 
carcinoma 
patients.  
\end{abstract}



\section{Introduction}

Data availability is a crucial issue that arises in the context of statistical model 
development and validation, inference derivation, introduction of statistical concepts
and many others \citep{Burton2006, Morris2019, DeBin2019}.  
In some  cases, especially for high-dimensional data (HDD) where the number of features 
is substantially larger than the number of observations, the number of data samples and 
data sets is not large enough to properly and reliably perform all required tasks. 
To overcome that deficiency, alternative data generation approaches such as the 
generation 
of artificial data are required. 
The generated data should match as close as possible the real-life data underlying the 
research question of interest, in particular with respect to its probabilistic structure 
as well as possible dependencies. 

A common approach for data generation is statistical data simulation. 
In this paper, we interpret data simulation as a data generation procedure that 
follows a data-generating process (DGP), with marginal distributions and a dependence 
structure as its basic components. 
For explanatory and prediction models being in focus of the present paper, data generation
procedures should necessarily include some steps on the generation of outcome 
variable(s). 
To this end, outcome-generating models (OGMs) are usually utilized to generate the 
outcomes 
using the available covariate information. 
Parameters of the OGM are to be estimated from real data, taken from literature or just
set based on investigator's choice. 
Examples for outcome data generation can be found in 
\citet{Reeb2013, Franklin2014, Binder2017, Atiquzzaman2020} and many others. 
\citet{Shmueli2010} depicts the application of OGMs both for explaining and 
prediction purposes.

For simulated data, the "truth" is assumed to be known a priori, at least to a some extent
\citep{Mehta2004}. 
That "truth" can be represented by simulation parameters or prespecified effect sizes, 
and  
is used to reliably evaluate the obtained estimates or predictions.

The most detailed practical introduction to simulation studies that includes structural 
approaches for their planning and reporting  as well as the discussion on the appropriate 
performance measures is presented in \citet{Morris2019}. 
In particular, the authors introduce a coherent terminology on simulation studies, 
data-generating mechanisms, and provide guidance on coding simulation studies. 
Most prominently, they introduce the {\bf A}ims, {\bf D}ata-generating mechanisms,
{\bf E}stimands, {\bf M}ethods, and {\bf P}erformance (ADEMP) criteria as a guidance 
for the planning and performing of simulation studies. 

For those researchers who use the results of simulation studies without being 
familiar with the entire simulation process, the discussion presented in 
\citet{Boulesteix2020} can be of great assistance. 
The paper not only contains many useful examples and applications, but also describes 
basic principles of simulation studies, gives insights into sampling variability and data 
generating processes, and demonstrates the role of statistical simulations in health 
research.

Two common approaches for data generation are parametric
and plasmode simulation, with the first approach being the most extensively 
studied and widely used one. 
Parametric simulations assume that the parametric stochastic model used to generate 
data is realistic and representative, with parameters of interest estimated from real data, 
derived from the literature or even set up by the user, in order to model 
specific scenarios \citep{Burton2006, Morris2019}. 
Plasmode data generation usually begins by resampling covariate information from the 
original real data \citep{Mehta2006, Gadbury2008}.
External (parametric) "truth" such as effect sizes or model parameter values 
in explanatory or prediction models can then be added to the covariable data sets to
define the relationship between the covariables and the outcome. 
With the OGM being a part of the plasmode data generation
procedure, the resulting plasmode simulation
can then be viewed as a semi-parametric data generation procedure. 
Of note, parametrically simulated data may often be considered to be purely artificial, 
whereas plasmode data is claimed to reflect reality in the most close way
\citep{Mehta2004, Mehta2006, Vaughan2009}.  

In the present paper, we provide an extensive literature overview of parametric and 
plasmode simulations. 
In particular, we discuss the difference between biological and 
\textit{statistical plasmodes}. 
We address advantages and challenges of parametric and statistical
plasmode simulation approaches in various contexts and provide step-by-step
recommendations for the generation of statistical plasmodes. 

At this point, we have no intention to demonstrate the 
superiority of a plasmode simulation over a parametric simulation or vice versa. 
We aim to analyze advantages and challenges of both data generation methods, 
and to illustrate the usefulness of plasmode simulations as a
complement to and possible extension of parametric simulation studies.

The paper is organized as follows: Section \ref{sec:paramplas} discusses parametric and 
plasmode simulations, compares their characteristics and provides an extensive literature 
review on both data generation methods. 
Section \ref{sec:pitfalls} analyzes statistical plasmodes in more detail by discussing 
their challenges. 
Section \ref{sec: recommendations} provides recommendations for planning, performing 
and reporting of statistical plasmode simulations. 
In Section \ref{sec: example}, we present a numerical example to 
illustrate the application of such a data generation approach. 
Section \ref{sec: disc} concludes with a discussion. 


\section{From Parametric to Plasmode Simulation Studies}
\label{sec:paramplas}

This section provides a comparative introduction to parametric and 
plasmode simulations. 
In particular, we start with a description of the main properties of 
parametric simulations as one of the most well-established types of data generation. 
Then we move on to plasmode simulations which are often claimed to be a more 
close-to-reality approach for data generation.

\subsection{Parametric Data Simulation}

In cases where an underlying DGP is defined in closed form and represented by a
parametric stochastic model, we speak of parametric simulations. 

The main asset of parametric simulations is their flexibility in terms of the chosen DGP. 
That is, one can easily generate a variety of independent data sets by varying the 
assumptions imposed on the DGP and its crucial parameters.
As a result, the simulated data sets can cover many complex but relevant 
scenarios as well as extreme situations that do not reflect reality. 
This feature becomes particularly important whenever we aim to analyze the behaviour 
and 
performance of different statistical methods. In addition, the knowledge of the DGP makes 
the corresponding parametric simulation more transparent and plausible.

Parametric simulations depreciate the sample size issue as an unlimited 
amount of data can be generated by means of a particular DGP. 
For instance, when applied to simulation of continuous random variables, an ``infinite'' 
number of distinct data points can be generated without much effort. 

The corresponding DGP represents a cornerstone of any parametric simulation. 
However, the existence of an appropriate model for the DGP that 
best fits some underlying real data cannot be taken for granted. 
On the other hand, even with a DGP available, the conclusions based on parametric 
simulations might be limited or even biased by the parameters of the chosen DGP model.

Obviously, the quality of a parametric simulation depends on the level of 
our comprehension of the underlying processes, distributions and possible dependencies. 
For instance, \citet{Vaughan2009} state that simulations 
might not fully reflect the complexity of the biological data that originates 
from nonrandom mating, recombination, hot spots, and other genetic mechanisms. 
\citet{Boulesteix2020} provides a similar statement and claims that many simulation 
studies are too simplified to describe the complexity of the real life data 
and thus may lead to inaccurate or even misleading findings.

When generating new data, we strive to preserve not only the marginal distributions
but also the underlying dependence structure.
In this context, the specification of appropriate dependence metrics, 
such as correlation, emerge.
One of the options for such a specification is estimation of the dependence 
structure from the data at hand. 
Computationally, such an estimation can be very expensive and time-consuming, 
especially in case of large data sets. 
For parametric simulations, this computational issue is  one of possible reasons for 
making independence assumptions on certain variables that then leads to a block diagonal 
structure for the corresponding correlation matrix \citep{Mehta2006}. 
Furthermore, a multivariate normal distribution provides the simplest model for 
multivariate covariate data with a pre-specified mean and correlation structure
\citep{Burton2006}. 
For generating non-normal correlated data, diverse copulas or the extended 
Fleischman power method can be utilized \citep{Headrick1999}.  
Obviously, certain concerns about the accuracy and realism of the underlying 
modelling  assumptions emerge for all such approaches. 

One of the assumptions imposed in the context of parametric simulations is that the 
"truth" 
must be known a priori. 
Such an assumption may not hold in both low- and high-dimensional situations. 
However, high-dimensionality may even exacerbate this issue  
making parametric simulations inapplicable. For instance, the “truth” about the set of 
biological markers truly associated with a given outcome may be unknown 
\citep{Azuero2012}. 

Within the framework of a parametric simulation, the underlying dependence structure
has to be completely specified in advance. 
The specification of such a dependence structure for 
high-dimensional data may become a challenge. 
Possible reasons can be not only computational efficiency issue, but also spurious 
correlations \citep{Fan2014}, sparsity of the data, some nonlinear or even hidden 
dependencies and other issues related to large covariance matrices; 
for more discussion and examples see \citet{Fan2006, Johnstone2009} and the 
references 
therein. 
Pitfalls in the specification of the dependence structure may then lead to false research 
discoveries and incorrect statistical inferences.

Dimensionality reduction and feature extraction play pivotal roles and are often 
fundamental 
in many high-dimensional settings \citep{Fan2006}.
However, it is not obvious how a parametric simulation may impact the findings of those 
procedures considering possible non-representativeness of parametrically generated 
covariate 
data sets.

Altogether, in many cases parametric simulations may turn out to be infeasible for 
high-dimensional data generation as it is not obvious how such simulations would cope 
with 
features of high-dimensionality.

A number of papers share our concerns on the applicability of parametric 
simulations for the generation of high-dimensional data sets. 
For instance, \citet{Gadbury2008} question the applicability of standard simulations 
performed in a high-dimensional experiment where hundreds of hypotheses are 
to be tested. 
Also \citet{Franklin2014} sees the application of ordinary simulation methods  
as an issue when comparing high-dimensional variable selection strategies.
In particular, the authors point out that the performance of those strategies depend 
"[\ldots]  on the information richness and complexity of the underlying empirical data 
source" and it is doubtful whether a parametric simulation is able to capture
the richness and complexity of the information. 

To overcome, at least partially, the limitations of parametric simulations outlined above,  
plasmode simulations have been introduced \citep{Mehta2006, Vaughan2009, 
Franklin2014}.

\subsection{Plasmode Data Simulation}

The term "plasmode" has been first introduced in \citet{Cattell1967}, with a plasmode 
data set defined as " [\ldots] a set of numerical values fitting a 
mathematico-theoretical model". 
In their seminal work, the authors emphasized that the certainty for the produced 
plasmode data set to fit the model comes either because there is a real life experiment 
producing the data of that kind or because the simulated data is produced mathematically 
to fit the functions.
Two different approaches for plasmode generation, performed either in a lab experiment 
or by resampling, were also mentioned in \citet{Mehta2006}. 
In the present paper, we refine the discussions provided by these authors, 
introduce the concept of statistical plasmodes and analyze their properties. 
From our perspective, the classification of plasmodes in biological and statistical 
depends on the procedure used for their generation.

Biological plasmodes are those generated " [\ldots] by natural biological processes, under 
experimental conditions that allow some aspects of the truth to be known" 
\citep{Vaughan2009}. Such plasmodes may be created, e.g., in a wet lab by manipulating 
biological samples as in case of a "spike in" experiment \citep{Mehta2006, Vaughan2009}. 
The latter paper provides a very illustrative introduction to biological plasmodes.  
In their detailed definition, the authors state that " [\ldots] a plasmode can be 
defined as a collection of data that 
(i) is the result of a real biological process and not merely the result of a computer 
simulation; (ii) has been constructed so that at least some aspect of the "truth" 
of the DGP is known”.

A number of research papers such as \citet{Mehta2004,  
	Vaughan2009} deal with "spike in" experiments in microarray expression 
analysis as an example for a biological plasmode data set. 
As part of that experiment, real cases from one population are randomly assigned to two 
groups. Then, a known amount of transcript is added to serve as a positive control. 
As a result, distributions and correlations in the generated data are viewed as most 
realistic since being taken directly from real data.
Besides others, \citet{Mehta2006} discusses application of plasmode data sets in high-
dimensional biology.
\citet{Vaughan2009} use  plasmodes for the estimation of admixture, or the 
proportion of an individual's genome that originated from different founding populations
and thus illustrates the utility of plasmodes in the evaluation of statistical genetics 
methodologies.
Several authors such as \citet{Sokal1980,  Mehta2006} provide helpful 
insights into generation and application of biological plasmodes as those to be
expected to incorporate valuable information on biological variation and capture biological 
reality.  
Biological plasmode data sets have also been utilized to evaluate the performance of 
statistical methods \citep{Irizarry2003} and their validity \citep{Mehta2004}. 
Plasmode data sets were also used to investigate the validity of 
multiple factor analysis in a known biological model \citep{Sokal1980}.

Despite their ability to create new and more advanced biological set-ups,  
e.\,g.\,, by crossing mice \citep{Vaughan2009}, sometimes biological
plasmodes may become not only very time-consuming but also require high experimental 
costs. 
Researchers might eventually do not have a lab available to construct biological 
plasmodes. 
In some cases, ethical reasons may also speak against the construction of biological 
plasmodes.
In all such situations, \textit{statistical plasmodes} offer an advantageous 
alternative. 

\textit{Statistical plasmodes}, being in focus of the present paper, begin with generation 
of covariate information performed  by applying resampling-based methods to real data set
\citep{Tibshirani2006, Reeb2013, Franklin2014}; 
note that no  biologically new samples are created in the context of such resampling 
procedure. 
Further, an appropriate OGM has to be applied to generate outcomes based on the 
resampled 
covariates \citep{Franklin2014, Franklin2015, Franklin2017, Karim2021, Desai2019,  
	Ripollone2019, Liu2019, Atiquzzaman2020, Conover2021, Wyss2021, Hafermann2022, 
	Rodriguez2022}. 
In cases where the exposure modeling is also a part of the study 
\citep{Franklin2014, Franklin2015, Franklin2017, Ripollone2019, Conover2021}, some 
known 
``truth'' such as e.\,g.\,treatment effects can be added manually 
\citep{Franklin2014, Liu2019, Desai2019, Atiquzzaman2020, Conover2021}.  
For instance, in \citet{Franklin2014} the authors create statistical 
plasmode data sets by " [\ldots] resampling from the observed covariate and exposure 
data 
without modification to preserve the empirical associations among the variables.” 
In that paper, the ``true'' treatment effect and the baseline hazard function 
are estimated from the empirical data. 
Further, the associations between outcome and the covariates as well as between 
censoring 
times and the covariates have been described by means of two Cox proportional 
hazard models. 
Such modeling approach corresponds to the application of a OGM in our 
terminology. 

According to our interpretation, statistical plasmode simulations utilize aspects of 
resampling (when generating the covariate information) as well as 
parametric modeling (e.g., application of OGM, modeling of exposure etc.) 
and thus can be interpreted as semi-parametric methods.

There are numerous applications of plasmodes generated by certain methods of data 
modifications 
in the literature.
Some of those applications are based on statistical approaches in the sense of our 
definition. 
For instance, \citet{Tibshirani2006} utilize plasmodes to assess sample size requirements 
in microarray experiments when estimating the false discovery rate and false negative rate 
for a list of genes. 
\citet{Gadbury2008} illustrates use of plasmodes by comparing the performance of 15 
statistical methods for estimating the false discovery rate in data from an high-dimensional 
experiment. 
\citet{Elobeid2009} employs plasmode data sets to analyze the performance of several 
statistical methods used to handle missing data in obesity randomized controlled trials.
\citet{Reeb2013} suggest an interesting application of plasmode data sets to complement 
the evaluation of statistical models for RNA-seq data. 
In their subsequent paper, \citet{Reeb2015} then use plasmode data sets to assess 
dissimilarity measures for sample-based hierarchical clustering of RNA sequencing data. 
In \citet{Franklin2014}, plasmode-based studies are used for the evaluation of 
pharmacoepidemiologic methods in complex healthcare databases. 
Resampling in combination with outcome generation by a logistic model to compare
the HDD propensity score method with ridge regression and lasso is used by 
\citet{Franklin2015}. 
\citet{Franklin2017} use plasmode-based studies to compare the performance of 
propensity 
score methods in the context of rare outcomes.
In \citet{Desai2019}, the authors utilize plasmode data sets to analyze the uncertainty 
in using bootstrap methods for propensity score estimation whereas 
\citet{Liu2019} conducts a plasmode-based study to compare the validity and precision of 
marginal structural models estimates using complete case analysis, multiple imputation, 
and inverse probability weighting in the presence of missing data on 
time-independent and time-varying confounders.  
The issue of data imputation has also been addressed in \citet{Atiquzzaman2020} where
the authors used plasmodes to compare two imputation techniques when imputing body 
mass 
index variable in osteoarthritis-cardiovascular disease relationship. 
In \citet{Ejima2020}, the authors use statistical plasmodes to
assess type I and type II error rates of analyses commonly used in murine genetic
models of obesity.
Similarly, \citet{Alfaras2021} resample from the empirical distributions to create 
plasmode data sets for murine aging data. 
Those plasmodes are then utilized to compute type I error rates and power for 
commonly used statistical tests without assuming a normal distribution of residuals. 
In their most recent study, \citet{Hafermann2022} designs a plasmode simulation study to 
investigate how random forest and machine learning methods may benefit from external 
information provided by prior variable selection studies. 
\citet{Rodriguez2022} evaluate plasmodes as being useful for preserving the underlying 
dependencies among hundreds of variables in real-world data used to evaluate the 
potential 
utility of novel risk prediction models in clinical practice; the authors generate plasmodes 
when studying  lung transplant referral decisions in cystic fibrosis. 

To our understanding, two central steps in the \textit{statistical 
	plasmode} generation procedure can be derived, namely: 
\vspace{-0.4cm}
\begin{itemize}
	\item [(i)]  {\bf Generation of the covariate structure} by resampling from an original data 
	set
	\item [(ii)] {\bf Outcome generation} that includes
	\begin{itemize}
		\item [(ii.1)] Choice of an appropriate outcome generating model (OGM)
		\item [(ii.2)] Choice of covariate effects either by individual specification or by 
		estimation based on the original data
		\item [(ii.3)] Generation of new outcomes by applying the  OGM chosen in (ii.1), with 
		the effects specified in (ii.2) and applied to the covariates  generated in (i)
	\end{itemize}
\end{itemize}

\vspace{-0.35cm}

The discussion performed in this section is summarized in 
Table \ref{t1} that provides a comparative summary of parametric 
and plasmode simulation studies. 
 \begin{center}
	\captionof{table}{Parametric simulations versus statistical plasmodes: Similarities and 
	differences}
	\label{t1}
	\vspace{0.15cm}
	\begin{tabularx}{0.97\textwidth} { 
			| >{\centering\arraybackslash}X
			| >{\raggedright\arraybackslash}X 
			| >{\raggedright\arraybackslash}X | } \hline
		{\bf Feature}&{\bf Parametric Simulations} &{\bf Statistical Plasmodes}\\
		\hline
		\hline
		Data-generating process (DGP)   & DGP is to be specified in advance    &No DGP 
		specification is required\\ \hline
		Outcome-generating model (OGM) &   Parameters of a chosen OGM to be estimated 
		from data or derived from literature or set manually & Parameters  of a chosen OGM 
		to be estimated from the original data or derived from literature or set manually  \\ 
		\hline
		Range of possible scenarios &Arbitrary scenarios, in particular, extreme and rare 
		scenarios, can be generated& Only reality bounded to the sample at hand can be 
		generated\\ \hline
		Knowledge of "truth" &"Truth" must be completely known in advance & At least some 
		"truth" such as effect sizes should be known a priori\\ \hline 
		Data availability and representativeness &Irrelevant for simulations based on  
		literature results or previous knowledge & Crucial, as the simulated data is always 
		limited to the sample at hand  \\ \hline
		Reality reflection & Parametric simulations may not be able to capture the complexity 
		of real life data  &Plasmodes are expected to resemble the reality in the most accurate 
		way\\
		\hline
		High-dimensional data simulations& Usually time - and cost - consuming.
		Latent dependencies may also become an issue  & Mostly straightforward, as no 
		estimation of distributions and/or dependencies is required\\
		\hline
		Small sample sizes& Essentially uncomplicated, but may become an issue in cases 
		when simulation parameters are to be estimated from the real data  at hand &Difficult 
		due to resampling \\
		\hline
		Dependence structure & Becomes a challenge with complex dependencies &No 
		modeling/estimation of dependence structure is required\\
		\hline
		
	\end{tabularx}
\end{center}
That discussion as well as the supporting literature imply that plasmodes
provide an attractive supplement to parametric simulations in data-based research. 
In particular, it is expected that plasmode datasets resemble the reality most closely, 
especially regarding dependency structures. 
In the following, we will analyze plasmodes in detail to examine their
strengths and weaknesses in more detail.

\section{Challenges of Statistical Plasmode Simulations}
\label{sec:pitfalls}

Simulations studies are, at least in the scope of the present work, designed to enable the 
practical analysis of statistical methods. 
To this end, data generation should satisfy several criteria, such as, amongst others, 
to provide the basis for subsequent undistorted model comparisons or to enable a specific 
covariate dependence structure. 
Constructing, reporting and comprehending a parametric simulation study is mostly
straightforward and transparent, as the resulting data
is artificially generated in a target-oriented way. 
Critical steps in the construction of parametric simulations include, 
for instance, the investigator's choice of the outcome-covariable association. 
While this ambiguity is shared by statistical plasmode simulations,
many of the properties of statistical plasmodes are typically less obvious 
and verifiable because statistical plasmodes are designed with the complex task to
mimic reality in the closest way while simultaneously specifying some aspects of the truth.
The main advantage of statistical plasmodes lies in their ability to generate data with 
specific distributions and dependence structures without the need for explicit assumptions.
The assumption that statistical plasmodes can faithfully generate data that closely 
resemble reality has rarely been questioned. 
In practice, however, this assumption can present challenges. 
For instance, the lack of statistical analyses or simulations to verify the preservation
of dependence structures can undermine the reliability of the generated data.
Consequently, the advantages attributed to statistical plasmodes can also transform 
into challenges. 
Further potentially critical steps in the construction of plasmode data include the 
representativeness of the underlying data and the choice of the resampling scheme.
Below, we theoretically discuss these potential pitfalls in more detail while 
also providing corresponding examples from literature.


\subsection{Resampling of Covariate Information}
\label{subsec: resamp}

In our concept of statistical plasmodes, the simulation is based
on the generation of covariable information by resampling from a real 
data set. 
This has the intention to preserve the characteristics of the original underlying 
dataset such as, amongst others, the number and type of covariables and the 
corresponding
dependence structure, see e.\,g.\,\citet{Franklin2015, Atiquzzaman2020, Conover2021}. 
This preservation is primarily achieved through the use of appropriate resampling 
techniques. 
Consequently, the applied resampling scheme, which consists of specifying the
number of generated datasets ($N$) and the resampling technique, has central
importance for the generated plasmode (covariable) datasets. 

Of note, while utilizing resampling, statistical plasmodes are
arguably even more complicated to analyze because of the additional artificial outcome 
generation. 
Consequently, not all established theoretical results concerning resampling might be 
transferable to the full plasmode dataset but only to the plasmode covariable datasets. 
We use the terminology resampling and bootstrap interchangeably and indicate the 
concrete resampling/bootstrapping technique if necessary. 
The analysis of resampling methods is almost exclusively formulated in terms of the 
asymptotic performance of the bootstrap distribution $L^*$ of an estimator
$T$ (e.\,g.\,variance, confidence interval) applied to the empirical 
distribution of the resampled data \citep[e.g.]{Bickel1997}. 
For statistical plasmode covariable datasets, the estimator $T$ could be, for example, 
some function of the covariance matrix of the covariables (preservation of correlation 
structure). 
When considering the statistical plasmode procedure as a whole, the estimator of interest
$T$ typically utilizes the artificial outcomes, e.\,g.\,$T$ could be the linear predictor 
in ridge regression when investigating its performance compared to other models. 
The resampling is said to "have worked", if $L^*$ convergences weakly to $L$ (the 
theoretical distribution of $T$) for increasing sample size $n$ of 
the underlying data \citep[e.g.]{Bickel2008}.  
Otherwise, one speaks of ``bootstrap failure''. 
In the following, we discuss the influence of the chosen resampling scheme on the 
generated 
plasmodes in more detail, focusing mainly on the preserveness of the covariable 
information. 

\textbf{Number of Plasmode Datasets $N$.}
The specification of the number of resampled plasmode datasets $N$ is often performed 
ad-hoc
and potentially leads to different answers to the same question, in particular if
$N$ is specified as too small \citep{Andrews2000}. 
In the framework of bootstrap tests, \citet{Davidson2000} propose a pretest 
procedure for choosing the number of bootstrap samples to minimize the loss of
power due to $N$ being finite. 
A more general, data-dependent three-step procedure is proposed by 
\citet{Andrews2000} 
who estimate $N$ to achieve a desired accuracy of the approximation of the
bootstrap to the ideal ($N \rightarrow \infty$) distribution of the estimator of
interest.  
However, to the best of our knowledge, there is no general guideline to theoretically 
specify the number $N$ of datasets to be generated in a data-independent way 
(i.\,e.\,without already performing the resampling scheme) such that 
asymptotic resampling results hold with sufficient accuracy. 
Moreover, existing results might not be valid for statistical plasmodes
due to the additional artificial outcome-generation procedure.

In the plasmode literature, $N=500$ \citep[e.g.]{Franklin2014, Desai2019, 
Atiquzzaman2020} 
and $N=1000$ \citep[e.g.]{Liu2019, Ripollone2019, Hafermann2022} seem to 
be popular ad-hoc choices. 
We have not seen any application where the choice of $N$ was explicitly justified
or the convergence or stability of the subsequent analyses applied to the plasmode data 
have 
been checked for increasing $N$. 
In summary, the number of datasets can be critical aspect in the generation of
\textit{statistical plasmodes}, in particular if convergence of $T$ is not reached. 
In Section \ref{sec: recommendations} we provide some recommendations for determining
$N$ which we further illustrate in Section \ref{sec: example}. 

\textbf{Resampling Technique.}
Resampling can be performed without replacement as in the $n$-over-$m$ bootstrap 
(subsampling with $m<n$) and sample-splitting (cross-validation) bootstrap. 
Subsampling draws from the data-generating process of the original data 
\citep{Politis1999} and has been shown to lead to consistent estimators under minimal 
conditions, see Theorem $1$ in \citet{Bickel1997}, as long as the subsampling 
size $m$ and the size of the original dataset $n$ are appropriately specified.
Alternatively, resampling with replacement such as the $n$-out-of-$n$ bootstrap
(also called nonparametric bootstrap) can be utilized. 
Resampling schemes based on drawing with replacement draw from empirical probability 
distribution derived from the underlying data \citep{Politis1999} and require additional 
assumptions for consistent estimation, but are more efficient if the bootstrap "works" 
\citep{Bickel1997}.
However, the nonparametric bootstrap can fail, for example when the limiting distribution 
of the estimator has discontinuities, when estimating extrema and when setting critical 
values for some test statistics
\citep{Bickel1997, Bickel2001, Andrews2010}.  
As a remedy, the $m$-out-of-$n$ bootstrap (sampling $m\leq n$ with replacement) has 
been 
introduced to prevent bootstrap failure while losing efficiency if the nonparametric
bootstrap was consistent. 
Sampling fewer than $n$ observations has since been treated as a ``cure-all'' method
(being asymptotically valid under weak assumptions and not failing)
which has been critically discussed, for instance, in \citet{Andrews2010}.
A comprehensive overview of resampling techniques is provided, e.\,g.\,, in 
\citet{Bickel1997}. 

For increasing sample size $n$, estimators based on subsampling and the 
$m$-out-of-$n$ bootstrap become more similar as the probability of repeating 
observations 
decreases. 
Note that, contrary to subsampling, resampling with replacement allows for $m=n$. 
The additional requirements for the consistency of sampling with replacement 
compared to sampling without replacement mainly state, informally speaking, that
the influence of tied observations on the bootstrap estimator should be small
\citep{Bickel2008}. 

In the majority of literature concerning plasmode generation, the $m$-out-of-$n$
bootstrap has been used \citep[e.g.]{Franklin2014, Karim2021, Wyss2021}, whereas
the nonparametric bootstrap has been used by \citet{Rodriguez2022}, 
the sample-split bootstrap by \citet{Gerard2020} and subsampling by
\citet{Hafermann2022}. 
In some publications we did not find indications whether resampling was performed
with or without replacement, e.\,g.\,in \citet{Ju2019, Ripollone2019}. 
All in total, the type of resampling technique influences asymptotic properties
of the covariables and hence of the plasmode datasets, effects whether the 
resampling ``has worked'' and consequently impacts subsequent analyses on the 
generated plasmodes datasets. 
However, to the best of our knowledge, we have not seen any application in the literature 
concerning plasmode data generation in which the choice of a particular resampling 
technique has been explicitly justified. 

\textbf{Resampling Size $m$.}
Using resampling with replacement of size $m$, with $m\rightarrow \infty$ 
and $m/n \rightarrow 0$, typically resolves failure of the $n$-out-of-$n$ bootstrap, 
but requires the specific choice of $m$ as a key issue \citep{Bickel2008}. 
An adaptive rule for the choice of $m$ for subsampling and the $m$-out-of-$n$ 
bootstrap in the case of independent observations has been proposed by 
\citet{Bickel2008} 
and is further illustrated in our example in Section \ref{sec: example}.
Informally speaking, if $m$ is in the right range of values, the 
bootstrap distributions of the estimator for the similar $m$'s are close to each other, 
indicating consistency of the estimator. 
The rule provides an adaptive estimator $m^*(n)$ and leads to optimal convergence
rates of the estimator irrespective whether the nonparametric bootstrap would work
in the example (then $m^*(n)/n \rightarrow 1$ as $n\rightarrow \infty$) or would fail 
(then $m^*(n)/n\rightarrow 0$).

To the best of our knowledge, only fixed resampling sizes $m$ have been chosen 
in the plasmode literature, and we have not observed any explicit justification of
the specific value of $m$. In other words, $m$ appeared to be chosen arbitrarily.
For instance, \citet{Hafermann2022} used a selection of $m's$ 
($250, 500, 1,000, 2,000, 4,000$) which are small compared to the number of 
observations $n = 198,895$ while \citet{Liu2019} chose $m=500$ for a dataset with
$n=646$ and \citet{Atiquzzaman2020} sampled $m=75,000$ out of $n= 84,452$. 
Interestingly, different authors used different values of $m$ which have been chosen 
without justification
for the same underlying dataset, as exemplified for the NSAID dataset with $n = 49,653$. 
While $m=30,000$ has been picked by \citet{Franklin2014} and \citet{Ripollone2019} used
a comparably large $m = 25,000$ as well, \citet{Ju2019} chose a much smaller value
in $m=1,000$ and \citet{Wyss2021} set $m = 10,000$. 
In summary, when applying subsampling or the $m$-out-of-$n$ bootstrap, the value
of $m$ matters for the consistency of $T$, and should be properly justified and 
adapted to the underlying data and estimator(s) of interest. 

\textbf{Covariable Dependence Structure and HDD.} 
Resampling the covariable information has, amongst others, the aim of
preserving the covariable dependence structure of the underlying dataset; 
see e.\,g.\,\citet{Franklin2014, Karim2021, Conover2021}. 
Under some assumptions such as i.\,i.\,d.\,observations and finite fourth moments of
the covariables, \citet{Beran1985} have shown that the resampled covariance matrix
converges to the original covariance matrix for the nonparametric bootstrap when $n$ 
increases and the number of covariables $p$ is fixed (i.\,e.\,most HDD situations excluded, 
see also below). 
However, for other resampling schemes similar results have, to the best of our knowledge, 
not been shown. 
For the $m$-out-of-$n$ bootstrap, the optimal $m$ could be estimated with the 
estimator $T$ specified to represent the covariable covariance matrix in order to 
investigate
and ensure that the resampling scheme works (at least for that certain aspect
of the data), see also our example in Section \ref{sec: example}. 
However, other aspects of the covariable information, such as extreme values, 
might be more important in some applications.  

A well-discussed issue, in particular in the context of HDD, is the occurrence 
of spurious correlations. 
Amongst others, \citet{Fan2014} have shown that sampling $p$ independent normal 
$n$-vectors leads to empirical covariance structures strongly deviating from a 
diagonal matrix, in particular if $p \gg n$. 
However, the risk of spurious correlations is not limited to parametric simulations.
An increasing number of covariates $p$ increases the risk that the underlying data 
sample suffers from spurious correlations, which may be propagated to 
the generated plasmode datasets by resampling. 
Further spurious correlation is likely to distort the empirical covariance structure
of the statistical plasmodes, leading to even stronger deviations to the
population covariance matrix. 

For a fixed number of covariables $p$, the bootstrap has been shown to work in linear
models if $p/n$ is small \citep{Bickel1983}. 
If the number of covariables growths with the number of observations, 
\citet{Mammen1993} 
have shown that the bootstrap works for effect estimates in high-dimensional linear 
models if $p(n)\rightarrow \infty$ and $p(n)/n \rightarrow 0$ as
$n\rightarrow \infty$, and \citet{ElKaroui2018} have shown that the confidence intervals of 
the pairs and residual bootstrap in linear models are too wide if $p(n)/n \rightarrow c, c\in 
\mathcal{R}$. 

In summary, the goal of preserving the covariable dependence structure can be used
to determine an optimal resampling scheme. 
In particular in high-dimensions, the covariance structure could, however, be distorted by 
spurious correlations and whether resampling and subsequently statistical plasmodes
work in these scenarios might require additional research. 

\subsection{Representativity of the Underlying Data Sample}
\label{subsec: rep}

One of the main assets attributed to statistical plasmode simulations 
is that they are expected to preserve the complex real-world data structure by resampling
the covariable information from a real dataset. 
Naturally, an appropriate representative dataset has to be available and constitutes the
basis for the entire plasmode simulation study. 
Parametric simulations, on the other hand, can be artificially constructed
without requiring representative data. 
The data sample is expected to represent the population of interest. This limits
the generalisability of the results of the analyses that the plasmode simulation study was
designed for, which has been acknowledged, amongst other,
by \citet{Franklin2014, Liu2019, Atiquzzaman2020}. 

The data sample should satisfy the assumptions of the applied resampling technique.
As a result, the choice of the resampling technique depends strongly on the underlying 
dataset at hand.
Standard resampling techniques, such as discussed above, assume that the observations 
are 
independent \citep[e.g.]{Bickel1997}. 
This assumption is violated, for instance, if the observations
show clusters, repeated measures, population structure or longitudinal measurements. 
In this context, more sophisticated resampling scheme including block-wise resampling
have to be applied, for which most of the asymptotic results are not explicitly formulated
\citep{Bickel2008}. 

Depending on the underlying data and the resampling scheme, the characteristics of the 
original dataset to be conserved might not be reflected by the generated data. 
This is acknowledged by \citet{Karim2021} who state that
``[\ldots] it is possible that important confounders in the empirical study might
not remain important in the plasmode samples''. 

In summary, the generated statistical plasmode datasets depend strongly on the 
representativeness of the underlying real data and are limited to the population 
represented by the data sample. 
The resampling scheme should be adaptive to the characteristics of the real data
such as population structure, which have to be identified and reported. 

\subsection{Investigator's Choice of the ``Truth''}
\label{subsec: truth}

The concept of plasmode simulations is mainly based on preserving
the complex but realistic structure of the underlying data while inserting some 
``truth'' by investigator's choice. 
These specification can be manifold in type, and potentially distort the 
real-word characteristics of the generated plasmode data. 

\textbf{Artificial Covariables.}
Additional to resampling covariable information, important covariables
such as exposure or treatment variables can be artificially created to model
some aspects of the simulation study. 
For instance, \citet{Franklin2017} and \citet{Conover2021} model a binary exposure
variable in relation to confounder variables via logistic regression, while
\citet{Rodriguez2022} simulate covariables at a later stage of a longitudinal study.  
Naturally, artificially generating covariables does not preserve the full 
real-world setting and should be performed with care. 

\textbf{Artificial Outcome Generation.}
With the covariable information generated (by resampling or artificially), 
corresponding artificial outcomes are created, in our concept of statistical 
plasmodes, according to some outcome-covariable association specified by the
investigator. 
A straightforward way to create a transparent association between 
resampled covariates and the artificial outcomes is to utilize regression 
models specified by the combination of a link function (type of OGM) 
and the linear predictor (effect structure). 

The OGM determines the type of the artificial outcome (e.\,g.\,binary, survival)
and strongly influences subsequent analyses on the generated data. 
For instance, if the aim of the study is the performance assessement
of several models, the model closest to the chosen type of OGM has an advantage
induced by the investigator, leading to potentially distorted comparisons.
Most commonly, logistic regression is used for binary outcomes 
\citep[e.g.]{Franklin2017, Karim2021, Atiquzzaman2020} and the  Cox model for
survival endpoints \citep[e.g.]{Franklin2014, Franklin2015, Desai2019}, 
whereas \citet{Rodriguez2022} apply an exponential survival model. 
Normal linear regression is used by \citet{Liu2019}. 

Besides the type of OGM, the determination of the effect structure of
the corresponding linear predictor is vital. 
Parts of the effect vector have been specified by literature review \citep{Conover2021},
by sampling from independent standard normal distributions \citep{Ju2019}, 
by estimation on the original dataset \citep{Liu2019, Wyss2021} 
or manually by investigator's choice \citep{Franklin2014, Franklin2015, Desai2019}. 
Some authors specify the treatment or exposure effect by hand while estimating
the confounder effects on the original data \citep[e.g.]{Karim2021}. 
Specifying the value of the effects of covariables might represent a strong
intervention in the generation process of a realistic dataset. 
Potential problems include the creation of artificial outcome-covariate associations
and invalidating or even nullifying existing ``real'' associations between 
the covariables and the novel, artificial outcomes, in particular if the effects are set 
manually by investigator's choice. 
If the effects are estimated, they depend on the underlying data sample and estimation 
uncertainty is ignored. 
Additionally, the specification of the type of OGM influences effect estimates and
subsequent analyses might become problematic. 
For example, an effect vector estimated by a sparse method will induce advantages
of sparse methods in model comparisons on the generated data. 

In summary, a crucial assumption for both parametric and plasmode simulations 
is that the chosen outcome generation reflects realistic, natural or 
biological associations between outcome and covariables. 
Whereas the choice of the OGM and the effect structure is a natural aspect of 
parametric simulation studies, plasmodes are often described as closely depicting 
reality. 
However, the outcome data in statistical plasmodes are also 
artificially created while inducing some investigator's choice ``truth''. 
These manipulations of the real-data are harder to assess and less transparent
than in parametric simulations studies as the structure of the data is typically
more complex. 
In the end, plasmode generation also leads, at least in part, to artificial data 
and constructed associations.

\section{Statistical Plasmodes: Step-by-step Recommendations} 
\label{sec: recommendations}

\begin{figure}[t!]
\centering
\includegraphics[width=0.97\textwidth]{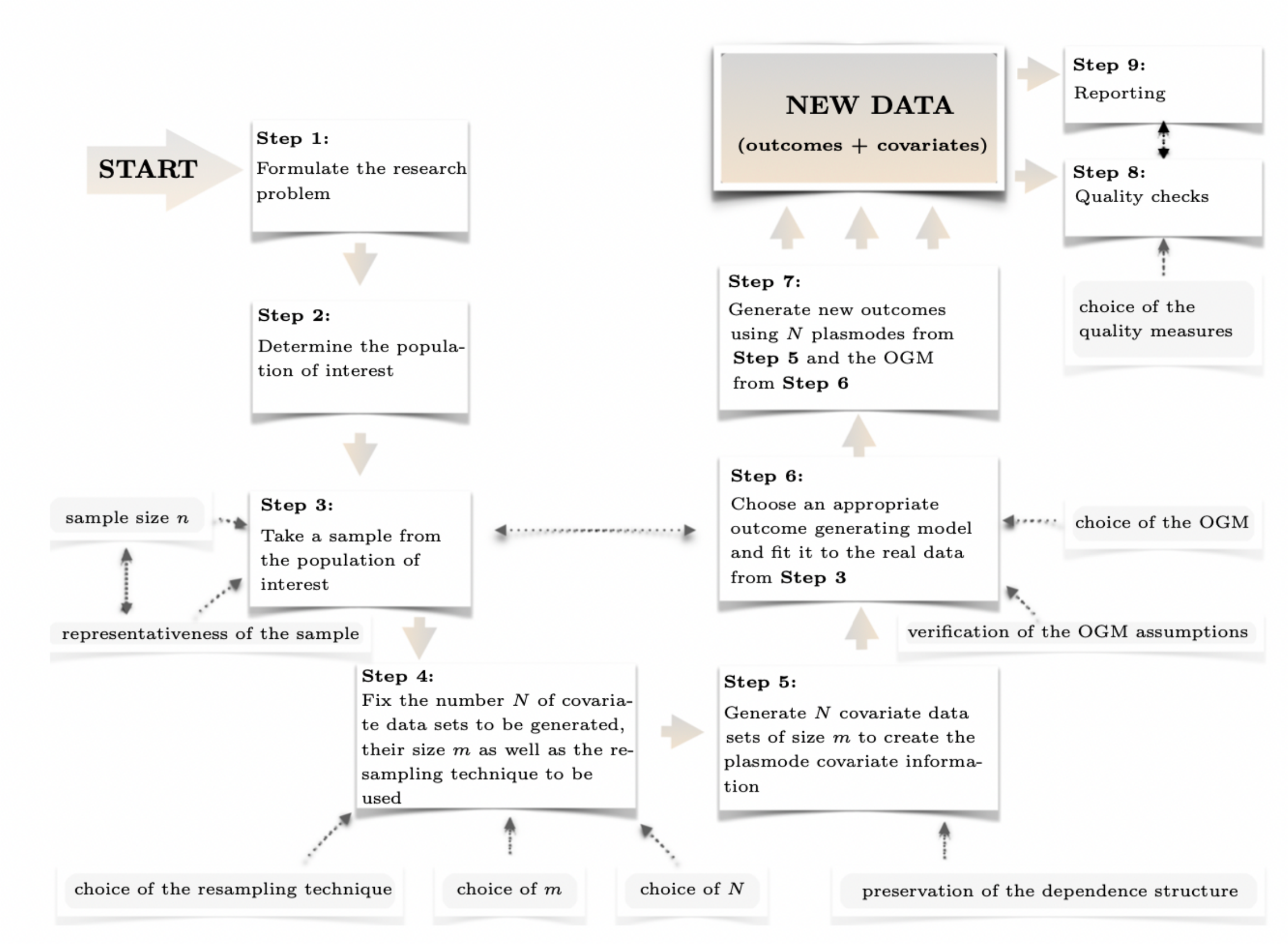}
\caption{Statistical plasmode data generation procedure step-by-step}
\label{fig:stepBYstep}
\end{figure}

We provide a hands-on overview of our recommendations for the generation and reporting 
of 
our concept of \textit{statistical plasmodes} in Figure \ref{fig:stepBYstep}.
This summary extends the basic plasmode generation procedure described in Section 
\ref{sec:paramplas} and addresses the critical steps discussed in the previous 
section. 
We theoretically discuss our step-by-step procedure below and illustrate its
application in a real data example in Section \ref{sec: example}. 

{\bf Step 1: Planning of the Simulation Study.} 
We recommend to clearly formulate the research problem and to plan the 
simulation study using the ADEMP criteria \citep{Morris2019}. 
Fixing the aims and the data-generating processes aids in the choice whether
statistical plasmodes are needed in the first place or whether a parametric 
simulation study might be more appropriate. 
Additionally, it guides the choice of the population of interest in Step 2 
and potentially the choice of the resampling technique in Step 4. 
Importantly, the methods and performance measures indicate the subsequent choice
of the OGM in Step 6. 
For instance, if we plan to assess the prediction performance of several models, 
we should make sure that the chosen OGM does not bias the subsequent model 
comparisons. 
Also, the choice of the OGM and data generating process determine the scenarios
in which the properties of novel statistical methods can be empirically assessed
in the context of the simulation study. 


{\bf Step 2: Population of Interest.} 
The population of interest can be of primary interest and consequently 
be strongly connected to the aim of the simulation study and the data-generating
processes determined in the previous step, in particular if methods are developed
to deal with populations with certain characteristics (e.\,g.\,many missing values, 
complex covariance structure, high-dimensionality). 
However, we might also be mainly be interested in the analysis of statistical methods
such that the population of interest serves mainly as an illustration and is not 
necessarily connected to Step 1. 
In the latter case, particular effort should be taken clarify why the chosen population
covers those situations in which the methods under consideration are claimed to work. 
In any case, the hypothetical population should be stated and described as clearly as 
possible, e.\,g.\,the entity of interest, covariables and population structure. 
Since the datasets generated by the statistical plasmodes should
be representative of the population of interest, this step influences several of
the following steps, in particular Steps 3, 4 and 6. 

{\bf Step 3: Representative Sample.} 
It is a central aspect of statistical plasmodes that the underlying sample
is representative of the population of interest clarified in the previous step, 
refer also to the discussion in Section \ref{subsec: rep}. 
Consequently, it is vital to investigate and communicate why the utilized 
data sample represents the population of interest and which potential limitations
arise. 
In particular, it should be stated how the data was sampled, which covariables 
are included and what the endpoint of interest is. 
Additionally, the sample size should be justified and potential population structures
investigated, as this can influence the choice of the resampling technique, see
Steps 4 and 5 and the discussion in Section \ref{subsec: resamp}.  
Note that even if the sample is representative, the generated plasmode datasets might
not be, e.\,g.\,as a result of a poor resampling plan or a outcome generation 
that distorts the relationship between the covariables or the outcome-covariable
association. 

{\bf Steps 4 \& 5: Resampling Scheme} 
The resampling scheme consists of the number of bootstrap samples, the
type of resampling technique used and, if applicable, the justification
of the resampling size, see also Section \ref{subsec: resamp}.
It determines, together with the data sample, the plasmode covariable datasets. 
Each of the aspects of the resampling scheme plays a crucial role for the asymptotic 
properties of the estimators applied to the data generated by resampling and to
properties such as the preserveness of the covariable correlation structure, 
see the discussions in Section \ref{subsec: resamp}. 
Unfortunately, the resampling scheme has to be decided on for each application
individually, while keeping those research aims and properties of the population 
of interest, that should be preserved with high priority, in mind. 
Ensuring the plasmode sets are drawn from the hypothetical population is only
possible by applying subsampling \citep{Politis1999}, whereas sampling with 
replacement draws from the empirical distribution of the data sample specified in
Step 3. 
However, if the underlying dataset is representative of the population 
of interest, drawing with replacement might become preferable due to its 
increased efficiency and second-order properties \citep{Bickel1997}. 
As described in Section \ref{subsec: resamp}, the nonparametric bootstrap 
potentially fails, although this is often impossible to know before the 
application. 
To avoid bootstrap failure, we recommend to utilize the $m$-out-of-$n$ bootstrap
although this might lead to efficiency losses. 
The optimal resampling size $m$ can be determined by applying the algorithm 
introduced in \citet{Bickel2008}
while using those properties of priority as estimator in Step 2 of the 
optimization algorithm for $m$. 
It has to be noted that the estimation of $m$ might require high additional
computational cost. 
In many applications it might be meaningful to opt for some function of 
the covariable covariance structure as an estimator, as it is often stressed that the 
empirical dependence structure of the original dataset should be preserved. 

{\bf Step 6: Outcome Generating Model.} 
The choice of the OGM includes the type of model to determine
the association between the resampled covariables and the novel artificial 
outcomes, as well as corresponding OGM components such as effect sizes, for example. 
The OGM determines the artificial outcomes in type and value, and is a crucial component
for many research questions formulated in Step 1.
Also, it gives the investigator 
the opportunity to fix some aspects of the ``truth'', see also the discussion in
Section \ref{subsec: truth}. 
Special care should be taken that the OGM does not bias the subsequent
analyses that the statistical plasmode simulations are generated for. 
To do so, it might be helpful to investigate the models or methods to be
compared in detail and contrast them with the OGM. 
For instance, a sparse OGM will most likely support sparse models in 
subsequent model comparisons. 
If the effect structure is chosen in a sparse way, a sparse model might
be more likely to correctly estimate the effect sizes or perform valid 
predictions. 
Additionally, if important relationships between variables have been detected,
the effect structure should be chosen accordingly to preserve these. 
For instance, in linear predictor models, the observed outcome variation
depends on the (co-)variances of the covariables weighted by their corresponding
effects, stressing their influence on the artifical outcomes
of the plasmode datasets.  

{\bf Step 7: Outcome Generation.} 
Each of the $N$ plasmode covariable sampled in Steps 4 and 5 is combined with 
the OGM determined in Step $6$ to create $N$ corresponding artificial plasmode
outcome vectors.

{\bf Step 8: Quality Checks.}
The quality of the covariables can be assured by appropriate resampling as
described in Steps 4 and 5. 
It is, however, often not feasible to compare the original covariable
covariance structure with those of the $N$ statistical plasmode
datasets. 
More research might be necessary to judge the distance of the original
and the generated data. 
The original outcome values of the real dataset are, if at all, only 
explicitly used to determine the effect structure in Step 6. 
The quality of the simulated data could be checked by comparing the
generated outcomes of (some of) the statistical plasmode datasets
with the original outcome. 
The type of potentially meaningful checks depends on the type of outcome. 
For continuous observations, the distributions of the two outcomes could be
compared by the empirical densities or histograms as is done e.\,g.\,in
\citet{Franklin2014}. 
Additionally, the range of the data should be checked as well as potential
outliers. 
For categorical (including binary) outcomes, the prevalence of the classes 
can be compared, see for example \citet{Franklin2014}. 

{\bf Step 9: Reporting.}
Each step of the statistical plasmode generation should be justified and
reported to enhance reproducibility and transparency of the proposed data 
generation procedure. 
Whenever appropriate, plasmode generation should follow the scheme presented 
in Figure \ref{fig:stepBYstep} and the corresponding descriptions provided in 
the present section. 
Additionally, the research question determined in Step 1 should be addressed.

\section{Statistical Plasmodes: A numerical example}
\label{sec: example}

The following example has been constructed to illustrate the step-by-step 
procedure introduced in the previous section. 
\newline

\begin{figure}[t!]
\centering
\includegraphics[width=0.97\textwidth]{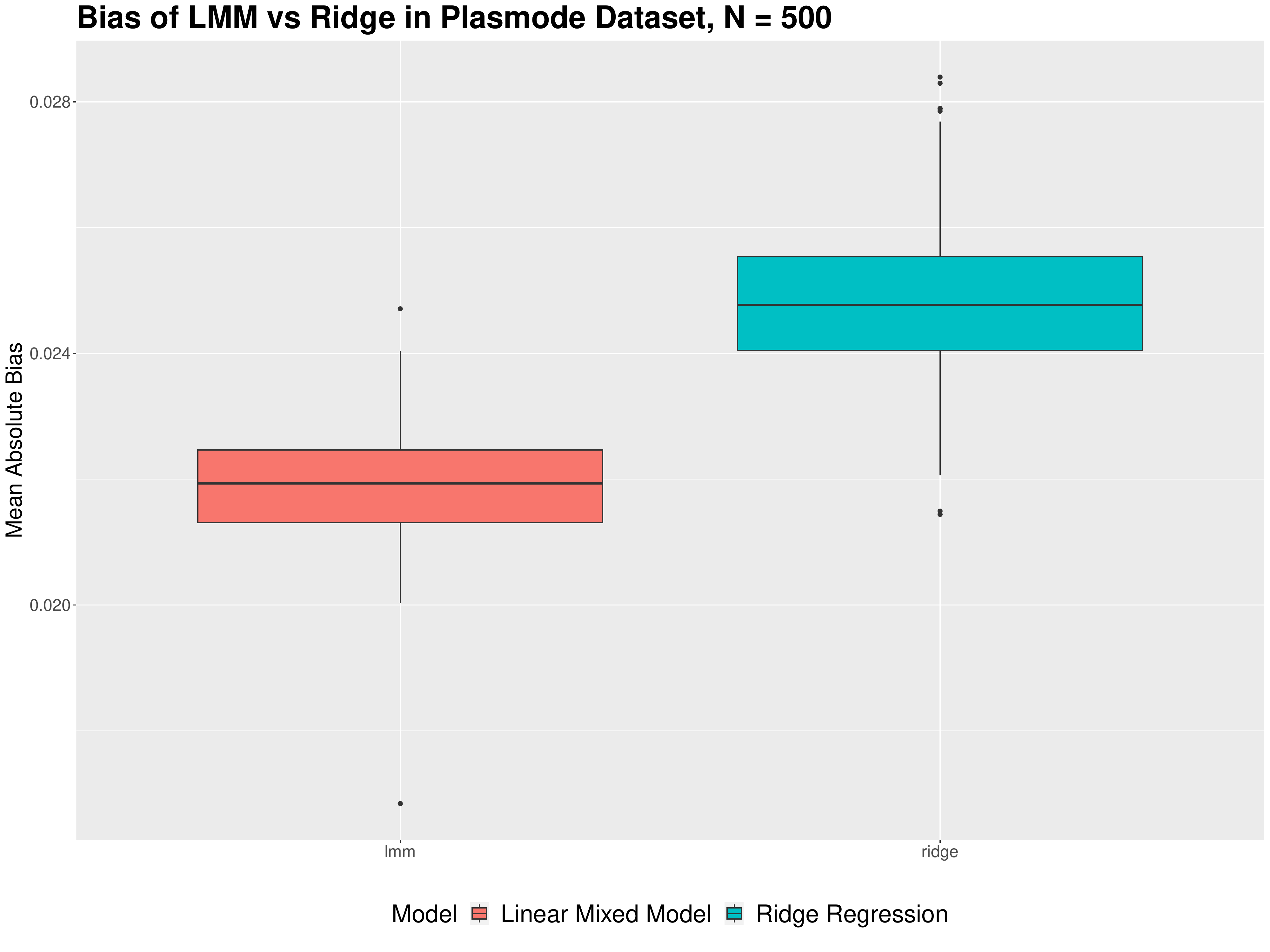}
\caption{Boxplots of the mean absolute bias for both the 
	linear mixed model and ridge regression in $500$ statistical plasmode datasets.}
\label{fig: bias}
\end{figure}

\begin{figure}[t!]
\centering
\includegraphics[width=0.97\textwidth]{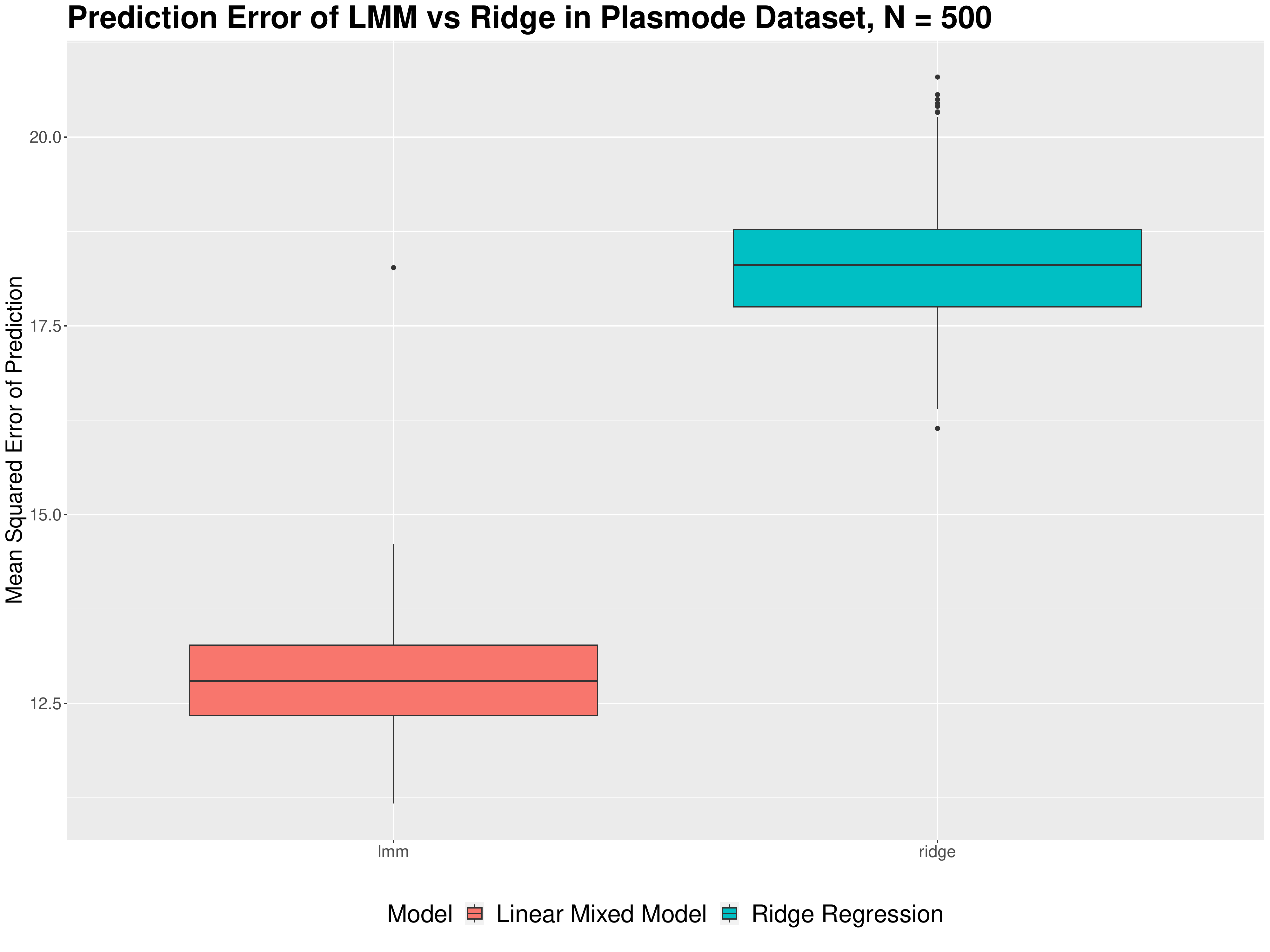}
\caption{Boxplots of the sample-split mean squared error of prediction for 
	both the linear mixed model and ridge regression in $500$ statistical 
	plasmode datasets.}
\label{fig: MSEP}
\end{figure}

\textbf{Step 1.}
Assume that we are interested in the aim (A) of investigating the application
of ridge regression \citep{Hoerl1970} and the linear mixed model \citep[e.g.]{Searle1992}
in the context of high-dimensional RNA-expression data with sparse effects on a normal 
outcome (data-generating process, D).
The estimands (E) are specified as the parameter vector and the linear predictor
in the respective model. 
We split the sample once into training and test data ($2:1$), which we deem sufficient 
for our illustration purposes. 
The plasmode datasets are generated using the training data. 
Each plasmode dataset of size $m$ and number of covariates $p$ is analysed (methods, 
M) 
using ridge regression of the form
\begin{align}
\ y = \mu 1_m + X\beta + \varepsilon, \quad \lVert\beta\rVert_{L_2}^2 \leq \lambda, \quad
\varepsilon \sim \mathcal{N}(0, \sigma^2 I_{m\times m})
\end{align}
via penalized maximum likelihood with cross-validation for $\lambda$ as implemented in 
the R-
package \texttt{glmnet} 
\citep{Friedman2010}, as well as the linear mixed model in the variance components form
\begin{align}
\ y = \mu 1_m + X\beta + \varepsilon, \quad \beta 
\sim \mathcal{N}(0, \sigma_\beta^2 I_{p \times p}), \quad \varepsilon 
\sim \mathcal{N}(0, \sigma_\varepsilon^2 I_{m\times m})
\end{align}
with restricted maximum likelihood estimation as implemented in the R-package 
\texttt{sommer} \citep{Pazar2017}. 
Here, $1_m$ denotes the $m$-column vector of ones while $I_{p\times p}$ denotes
the identity matrix of dimension $p$. 
As performance measures, we utilize the mean absolute bias
\begin{align}
\ \mathrm{MAB} = \frac{1}{p+1}\lVert (\hat{\mu}, \hat{\beta}) - 
(\mu, \beta) \rVert_{L_1}
\end{align}
where $\mu$ and $\beta$ are known as part of the ``truth'', 
and the sample-split mean squared error of prediction 
\begin{align}
\ \mathrm{MSEP} = \frac{1}{m}\lVert\hat{y} - y_{\text{test}} \rVert_{L_2}^2, 
\quad \hat{y} = \hat{\mu}1_m + X_{\text{test}}
\hat{\beta}
\end{align}
where $y$ corresponds to the artificial outcome in the test split. 
We estimate both measures using the mean of the estimates
(indexed by superscript $b$) in the generated $N$ statistical plasmode datasets 
\begin{align} \label{equa: perf}
\ \widehat{\mathrm{MAB}} =  \frac{1}{N} 
\sum_{b=1}^N \frac{1}{p+1} \lVert (\hat{\mu}, \hat{\beta} )^{(b)} - (\mu, \beta)
\rVert_{L_1}, \quad
\widehat{\mathrm{MSEP}} = \frac{1}{N} \sum_{b=1}^N \frac{1}{m}\lVert\hat{y}^{(b)} -
y_{\text{test}} \rVert_{L_2}^2.
\end{align}
and visualize the $N$ individual measures via boxplots, see Step 9 and 
Figures \ref{fig: bias} and 
\ref{fig: MSEP}. 

\textbf{Step 2.}
In the scope of this example, we are interested in the model choice for high-dimensional 
RNA-expression data with normal outcomes for female breast cancer patients which 
constitutes the population of interest. 

\textbf{Step 3.}
The data sample underlying the statistical plasmode simulation was generated by 
The Cancer Genome Atlas (TCGA) Research Network (\url{https://www.cancer.gov/tcga}).
The breast carcinoma (BRCA) cohort which provides a basis for the following numerical 
example was last updated on May 31, 2016. 

\begin{figure}[t!]
\centering
\includegraphics[width=0.97\textwidth]{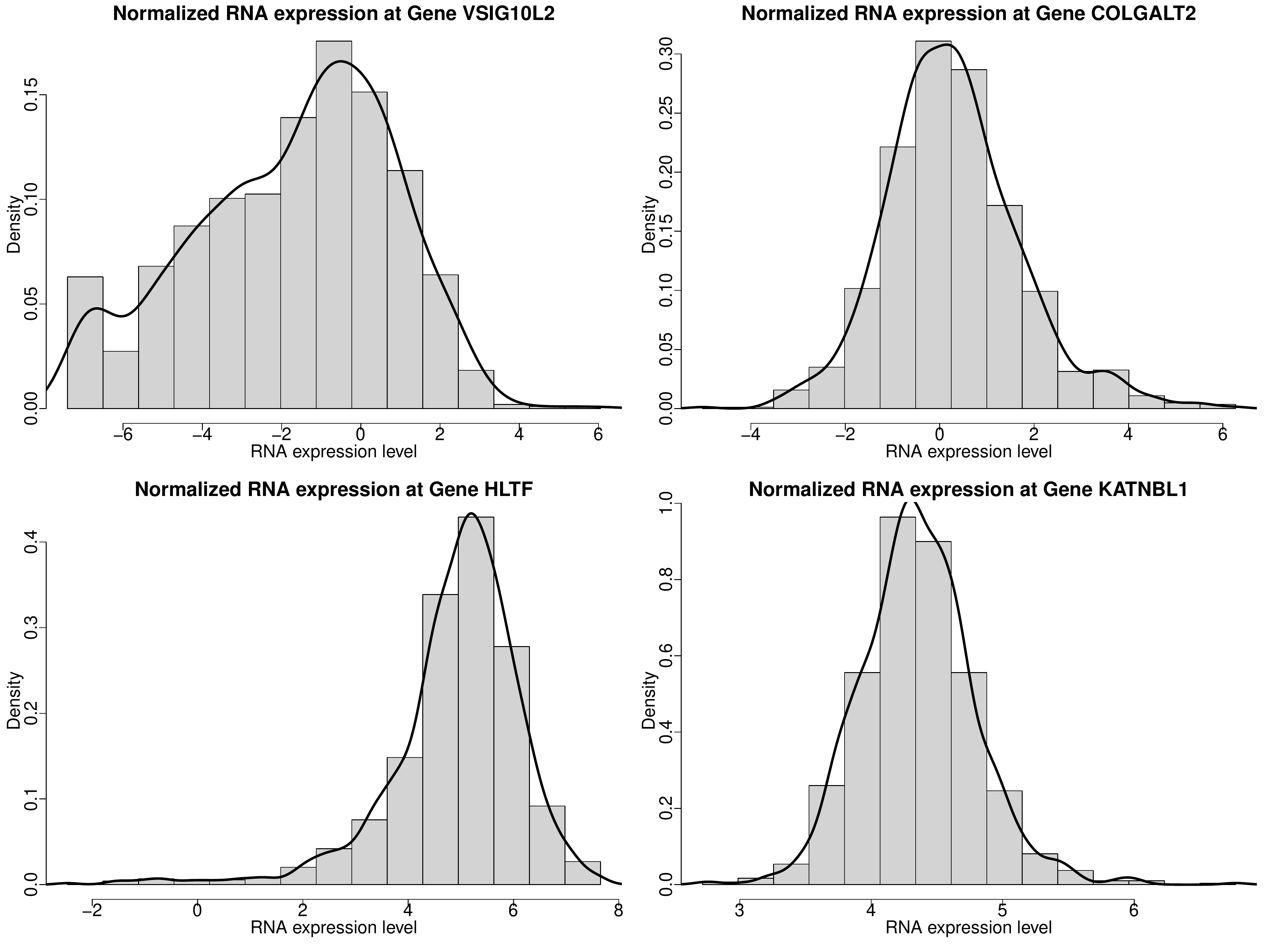}
\caption{Empirical distributions illustrated by histograms (15 breaks each) and
	smoothed densities for four genes selected at random to illustrate the differences
	in location and shape.}
\label{fig: RNA}
\end{figure}

We restrict the publicly available data to $n=1098$ female patients with breast 
cancer with cancer tissue, i.\,e.\,excluding normal tissue and male patients. 
RNAseqV2 gene expression data and clinical data for BRCA were 
obtained from the TCGA Data Portal 
\citep{Weinstein2013} via the R/Bioconductor package \texttt{TCGAbiolinks} 
\citep{Colaprico2015, Silva2016, Mounir2019}. 
For computational reasons, we choose $p=5000$ out of the $25828$ available genes 
at random. 
The R/Bioconductor package \texttt{limma} \citep{Ritchie2015} has been utilized to 
normalize the RNA gene expression data. 
The expression levels can be assumed to be measured continuously and they show 
different shapes and ranges. 
This is illustrated in Figure \ref{fig: RNA} using their empirical distributions 
at four randomly chosen genes.  

\begin{figure}[t!]
\centering
\includegraphics[width=0.97\textwidth]{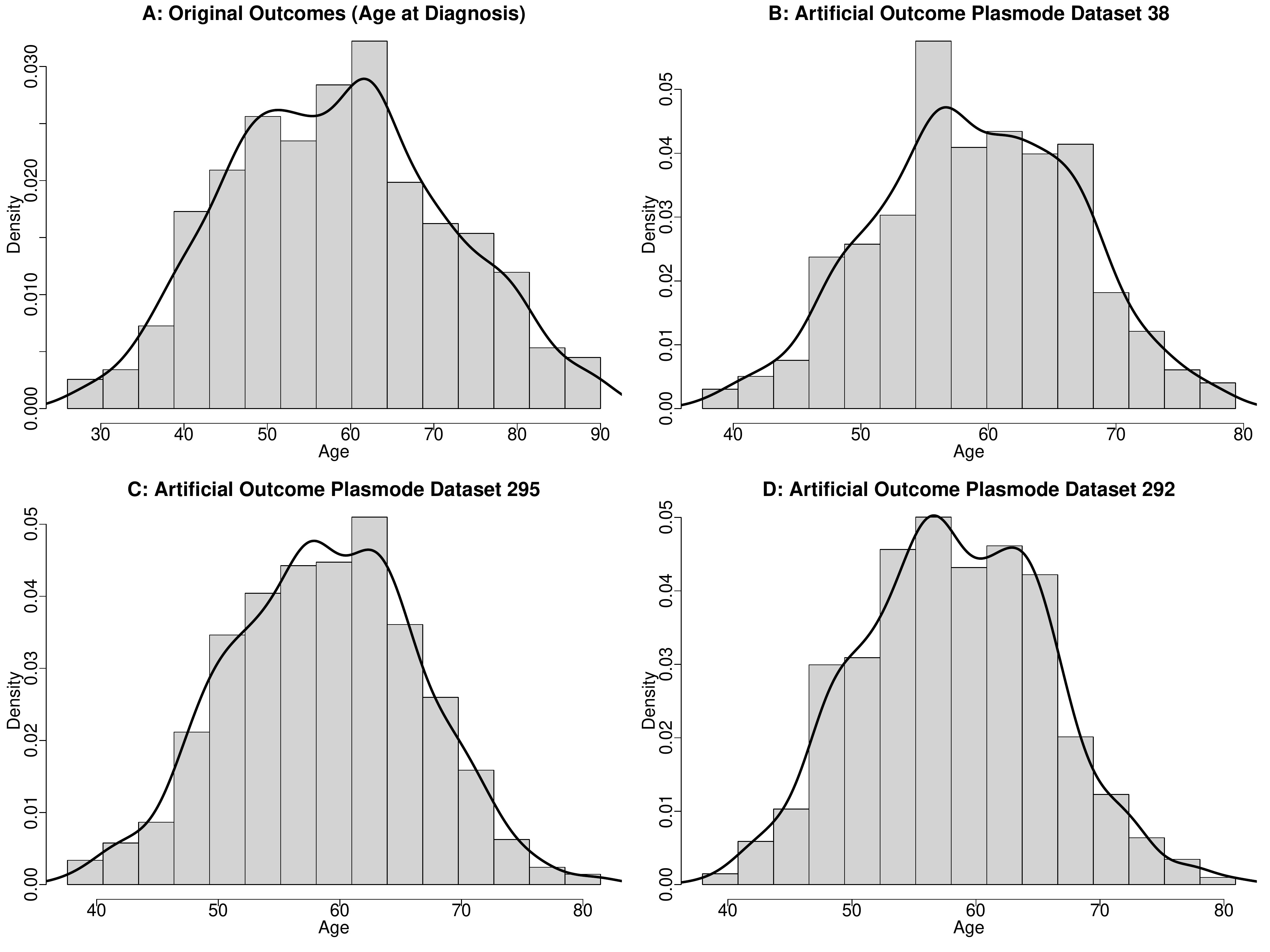}
\caption{ Empirical distributions illustrated by histograms (15 breaks each) and
	smoothed densities for 
	A: the original outcome (age at diagnosis) 
	and B-D: vs. artificial outcomes of three plasmode datasets selected at random. }
\label{fig: Outcome}
\end{figure}

The outcome of interest is age at diagnosis date which can be considered to be 
approximately normally distributed, see Figure \ref{fig: Outcome}A.  
While the dataset can be considered to be representative of a female breast cancer 
population from the United States of America, we acknowledge that RNA expression data 
from other populations (e.\,g\,.different countries) might lead to different results 
for our research question. 

\begin{figure}[t!]
\centering
\includegraphics[width=0.97\textwidth]{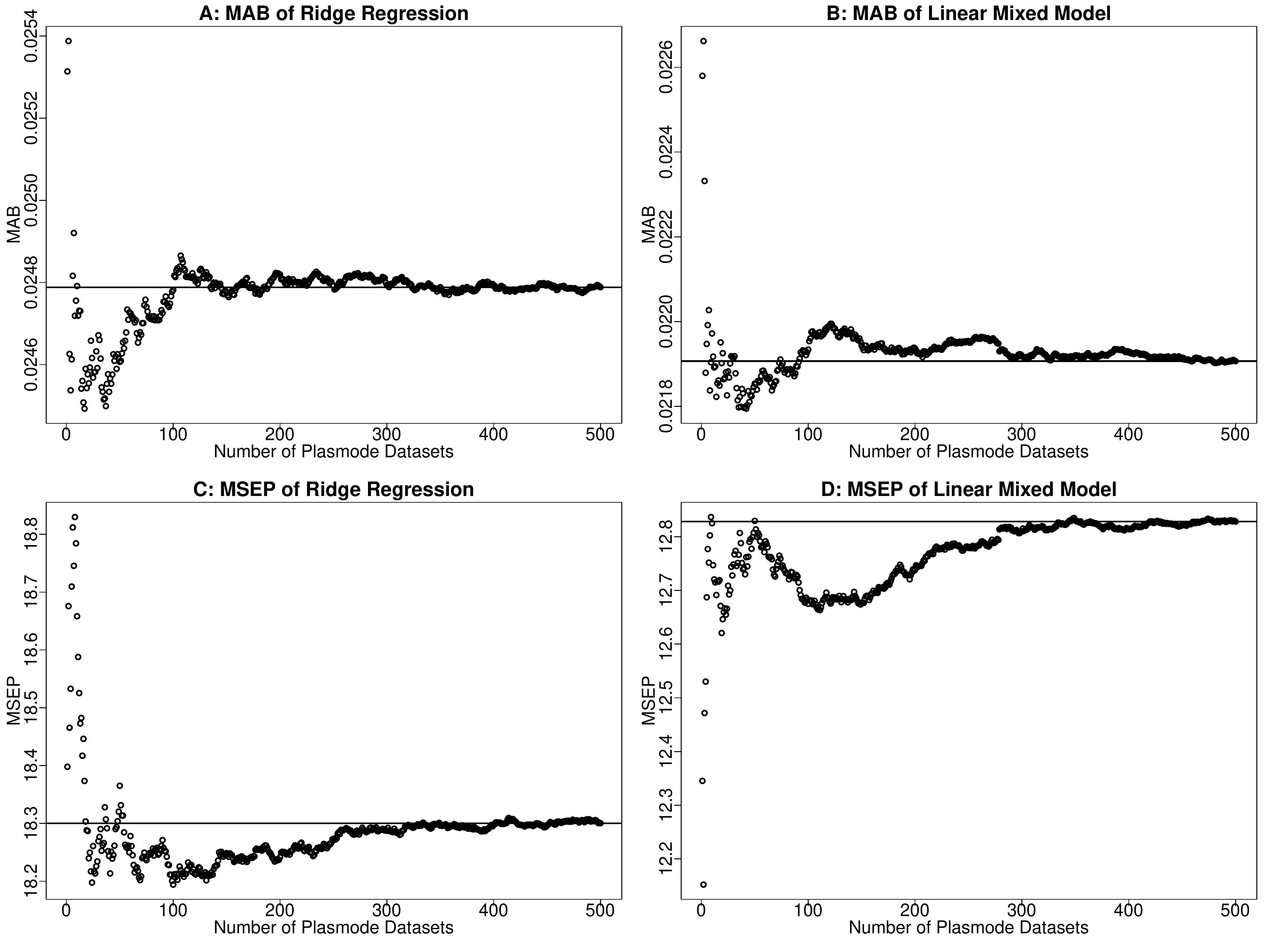}
\caption{Convergence of the performance measures MAB and MSEP for increasing 
	number $N$ of statistical plasmode datasets.}
\label{fig: conv}
\end{figure}

\textbf{Step 4.} 
Before the analysis, we set the number of plasmode datasets to be generated 
to $N = 500$. 
In the final step 9, we investigate the convergence of the estimators of the
performance measures, see equation (\ref{equa: perf}), in the
statistical plasmode datasets, see also Figure \ref{fig: conv}.

We choose the $m$-out-of-$n$ bootstrap in order to prevent potential bootstrap
failure but with the potential drawback of losing estimation efficiency. 
Performance analysis of resampling method and the estimation of the optimal $m$ 
requires the specification of an estimator which is applied to the generated data. 
Since resampling in statistical plasmodes is primarily concerned with the covariate 
information, already using the performance measures defined in equation 
(\ref{equa: perf}) as estimators is not feasible as they require the subsequent 
artificial outcome generation. 
Naturally, there are several reasonable estimators that could be considered. 
In this example, we opt for the covariate dependence structure as the measure of interest
because the majority of publications which applied plasmodes referred to the 
advantage of the preserveness of the original covariable dependence structure. 

We determine the resampling size $m$ via the algorithm described in 
\citet{Bickel2008}. 
In particular, to adopt that algorithm to our problem formulation, 
we specify the sequence of potential $m$'s by setting $q = 0.97$, choose
the $L_2$-norm of the covariance matrix (of the resampled covariate data) as a metric, 
and calculate the resulting empirical distribution functions.
We estimate the covariance matrix using the Ledoit-Wolf linear shrinkage estimator
\citep{Ledoit2004} to obtain a more precise estimate which is necessary because 
the covariate data are high-dimensional. 
The optimal resampling size $m^*$ is the one which minimizes the distance between 
the distributions of subsequent $m$'s, where the distance is exemplarily measured by 
the Wasserstein metric. 
The optimal resampling size based on the Wasserstein metric using $100$ iterations
resulted in $m^* = 711$. 

We acknowledge that there is variety of optimal resampling sizes $m^*$ if 
any of the parameters of the algorithms would be changed (such as, amongst others,
estimator, distance metric for empirical distributions and sequence of potential m's). 

\textbf{Step 5.}
We apply resampling with replacement of size $m^*= 711$ to the matrix of
covariable information to obtain $N = 500$ statistical plasmode covariable datasets. 
As we have determined the subsampling size with optimality criterion as the 
$L_2$-norm of the covariance matrix of the covariables, the empirical  
covariance structure of the original dataset should be sufficiently preserved. 

\textbf{Step 6.}
We choose the LASSO \citep{Tibshirani1996} as an appropriate OGM to represent the 
sparse
effect structure associated with the high-dimensional data as required in Step 1. 
Additionally, the LASSO most likely does not distort the comparison between
ridge regression and the linear mixed model as both of these methods 
are shrinkage methods used to model polygenic effects. 
The ``true'' effect structure for the LASSO is chosen as the vector of
estimated effect sizes obtained after a LASSO had been fit to the original data. 
The proportion of non-zero estimated effects was $95.4 \%$ ($4768$ vs $232$).  
This implies that $232$ covariables are selected in the investigator's choice ``truth''
while $4768$ genes are given a null effect. 
The effect sizes of the selected covariables has a median of $-0.01$
(range $[-2.97, 2.05]$). 

\textbf{Step 7.}
We generate one artificial outcome vector of size $m^*= 711$ for each of the 
$N=500$ plasmode covariate datasets by calculating the linear predictor based 
on the combination of the resampled covariable information (step 5) and the 
``true'' effects (step 6). 
Thus, we obtain statistical plasmode simulations based on real covariate information
with an investigor's choice ``truth''. 

\textbf{Step 8.}
The artificial outcomes (of some of) the $N=500$ plasmode datasets 
are compared with the original outcomes via histograms in Figure \ref{fig: Outcome}.  
The distribution of the original and artificial outcomes is very similar in
shape and mean.  
The range of the original outcomes is larger than the range of the artificial 
outcome which can be explained by the outcome generation via resampled covariables
and effects determined by LASSO (sparse and shrunken effects) which most likely will 
not lead to more extreme outcome values than contained in the underlying dataset. 

We conclude that the artificial outcome data come close to reality but might
not properly reflect extreme values. 
The range of the artificial outcomes could be increased, e.\,g.\,, by manually
altering some elements of the effect vector estimated by LASSO (as investigator's 
choice of the ``truth''). 
By doing so, however, we would further alter the association between some
of the covariables and the novel outcomes. 

\textbf{Step 9.}
Finally, we compare the performance of ridge regression and the linear mixed model
in our statistical plasmode simulations. 
The MAB of the Ridge regression is estimated as $0.025$ while the estimate of 
the MAB of the linear mixed model is $0.022$, see also equation (\ref{equa: perf}). 
The sample-split MSEP of ridge regression is estimated as $18.30$ while 
the sample-split MSEP of the linear mixed model is $12.83$, see also equation 
(\ref{equa: perf}).
In Figure \ref{fig: bias} and \ref{fig: MSEP} we depict the estimated values
for each plasmode dataset via boxplots. 
This suggests that in our generated statistical plasmode simulations, which 
represent high-dimensional RNA expression data with sparse known effects 
and artificial normal outcomes, the linear mixed model performs superior
to ridge regression. 

Additionally, we illustrate in Figure \ref{fig: conv} the convergence of the
performance measures for increasing number of plasmode dataset. 
The estimators for MAB and MSEP for both ridge regression and the linear mixed
model seem to have stabilized at about $300$ generated simulations. 
Thus, we conclude that the generated number of statistical plasmode is sufficient
to obtain stable estimates of the performance measures defined in step 1. 

\section{Conclusions and Outlook}
\label{sec: disc}

Many simulation studies impose relatively strong assumptions regarding the nature of 
randomness in the data and its dependence structure. 
Mostly of theoretical kind, those assumptions primarily rely on the assumptions inherent
in the statistical models applied to generate the data. 
Since not all assumptions can be justified in applied settings, the 
corresponding simulation studies may not be able to capture biologically meaningful 
relationships and thus result in misleading conclusions and research findings.

To avoid (at least some of) those issues, plasmode data sets are considered as an 
alternative data generation approach.  
While parametric simulations are known to provide only a partial representation of reality 
\citep{Reeb2013}, plasmodes have been declared to generate data that resemble reality in
the closest way \citep[e.g.]{Mehta2004}. 
Highly appreciated for their ability to generate most realistic data, plasmode do not impose 
any specific model assumptions on their data generation process. 
Thus, no assumptions need to be justified to address the applicability of plasmodes. 
Nevertheless, a number of assumptions such as the representativeness of the underlying 
data 
sample have to be verified in order to guarantee the reliability of the generated plasmode 
data.  

Plasmodes can accommodate unknown features such as dependence structure, 
distributions, 
and others, in particular, in the case of high-dimensional data. 
We recall that in case of parametric simulations most of those quantities are to be
specified in advance. 
All in total, plasmode data sets may provide an attractive supplement to parametric 
simulations and can  be applied in order to increase the reliability of the obtained
research results.

In  the present paper, we first discuss the concept of statistical plasmodes as those 
created by resampling of covariate information from empirical data at hand and 
subsequent 
outcome generation using an appropriate outcome-generating model. 
This is what distinguishes them from biological plasmodes  which are usually created by 
conducting lab experiments. 
We interpret statistical plasmodes as an intermediate step between the parametric and 
nonparametric simulations, with the parametric component represented by the chosen 
outcome-
generating model. 
After the introduction of statistical plasmodes, we discuss their main advantages and 
challenges and propose a step-by-step scheme for their generation and reporting.
That scheme is then illustrated by means of a numerical example. 
All discussions in the present paper are presented in the context of prediction and 
explanatory models.

Plasmodes are bounded to the sample they are based on, and thus cannot produce the 
same 
variety of different scenarios as parametric simulations do. 
In this context, questions on the data availability and representativeness arise. 
In particular, even if plasmodes offer a flexible data generation procedure which creates 
realistic data, the representativeness of the generated data still substantially depends on
the representativeness of the underlying real data set.
To address this limitation, some authors such as \citet{Ejima2020} assume that the 
empirical data at hand represents the entire population of interest. 
Of course, such an assumption cannot be satisfied in each particular situation.

Spurious correlations are another issue closely related to the question of 
representativeness. 
Although plasmodes do not specify the underlying dependence structure explicitly, they 
do 
reproduce it to a certain extent while generating new data. 
Thus, if the sample at hand does not adequately represent the population of interest, the 
existing spurious correlations may be increased or even distorted for the generated 
plasmode 
data sets. 
As a result, the corresponding generated dependence structure will not represent the real
one. 

Statistical plasmodes as introduced in the present paper incorporate features from both 
parametric simulations and resampling approaches, and, as a result, inherit the strengths
and weaknesses of each data generation method. 
On one hand, statistical plasmodes offer the advantage of creating more realistic data by 
generating covariate information through resampling techniques.
On the other hand, they may also introduce certain challenges with respect to the 
subsequent model comparisons, as compared to purely parametric simulations 
\citep{Wyss2021}. 
Statistical plasmodes enable to control and manipulate certain aspects of the ``truth"
through the use of parametric OGMs, which can be advantageous over pure resampling 
methods. 
Nevertheless, asymptotic results established for resampling techniques may not be 
directly 
applicable to statistical plasmodes.

Our discussion  points out several interesting options for future research. 
First, basic expectations placed on the plasmodes are related to their ability to  
preserve real data distributions, the underlying dependence structure and, as a result, the 
existing empirical associations. 
Those expectations  are to be guaranteed  by resampling from the observed covariate 
data at 
hand, without any additional data modification. 
However, it is not obvious how the choice of a particular resampling technique and 
specification of its parameters (such as the subsampling proportion in case of the 
subsampling technique) might impact the robustness of the obtained data generation 
results, 
e.\,g.\,, in the context of spurious correlations or sparse data.
Additionally, the calculation of the optimal $m$ might require high computational costs. 
A closer analysis of these impacts are possible topics for future research.

Second, a data generation method is considered to be realistic if it reflects the real data 
structure and the existing dependencies in the most accurate way.  
Thus, appropriate distance measures need to be specified in advance and also included 
into 
the reporting step of the data generation procedure. 
Such measures can then be used to measure the closeness of the generated plasmode 
data set 
to the underlying real data set. 
The choice of an appropriate distance measure, as well as the robustness of the plamode 
generation procedure with respect to that choice, can also be an interesting research topic.

Finally, the outcome-generating models present the major obstacle for plasmodes to 
become a 
purely non-parametric data generation approach. 
In the future we intend to analyze the impact of an OGM on the performance of the 
plasmode 
data generation procedure and to construct examples where the replacement of a 
parametric 
OGM with a non-parametric one improves the obtained data generation results. 
It is also of great interest to address possible ``plasmode failure" for data sets generated 
through statistical plasmodes.

In total, our paper presents a comprehensive analysis of statistical plasmode 
simulations, discusses their potentials and central challenges and provides
step-by-step recommendations for their generation. 
Our future research aims to address (at least some of) the pitfalls in the most close way to 
potentially provide more understanding and further novel insights into 
statistical plasmode generation.

\section*{Acknowledgements}
The authors would like to thank Jörg Rahnenführer, Andrea Bommert and Marieke Stolte 
for helpful discussions.

\bibliography{BibliographySiM2}

\begin{thebibliography}{59}
\providecommand{\natexlab}[1]{#1}
\providecommand{\url}[1]{\texttt{#1}}
\expandafter\ifx\csname urlstyle\endcsname\relax
  \providecommand{\doi}[1]{doi: #1}\else
  \providecommand{\doi}{doi: \begingroup \urlstyle{rm}\Url}\fi

\bibitem[Burton et~al.(2006)Burton, Altman, Royston, and Holder]{Burton2006}
A.~Burton, D.~G. Altman, P.~Royston, and R.~L. Holder.
\newblock The design of simulation studies in medical statistics.
\newblock \emph{Statistics in Medicine}, 25:\penalty0 4279--4292, 2006.

\bibitem[Morris et~al.(2019)Morris, White, and Crowther]{Morris2019}
T.~P. Morris, I.~R. White, and M.~J. Crowther.
\newblock Using simulation studies to evaluate statistical methods.
\newblock \emph{Statistics in Medicine}, 38:\penalty0 2074--2102, 2019.

\bibitem[Bin et~al.(2019)Bin, Boulesteix, Benner, Becker, and
  Sauerbrei]{DeBin2019}
R.~D. Bin, A.-L. Boulesteix, A.~Benner, N.~Becker, and W.~Sauerbrei.
\newblock Combining clinical and molecular data in regression prediction
  models: insights from a simulation study.
\newblock \emph{Briefings in Bioinformatics}, 00\penalty0 ((0)):\penalty0
  1--17, 2019.

\bibitem[Reeb and Steibel(2013)]{Reeb2013}
P.~D. Reeb and J.~P. Steibel.
\newblock Evaluating statistical analysis models for rna sequencing
  experiments.
\newblock \emph{frontiers in Genetics}, 4:\penalty0 1--9, 2013.

\bibitem[Franklin et~al.(2014)Franklin, Schneeweiss, Polinski, and
  Rassen]{Franklin2014}
J.~M. Franklin, S.~Schneeweiss, J.~M. Polinski, and J.~A. Rassen.
\newblock Plasmode simulation for the evaluation of pharmacoepidemiologic
  methods in complex healthcare databases.
\newblock \emph{Computational Statistics and Data Analysis}, pages 219--226,
  2014.

\bibitem[Schulz et~al.(2017)Schulz, Zöller, Nickels, Beutel, Blettner, Wild,
  and Binder]{Binder2017}
A.~Schulz, D.~Zöller, S.~Nickels, M.~E. Beutel, M.~Blettner, P.~S. Wild, and
  H.~Binder.
\newblock Simulation of complex data structures for planning of studies with
  focus on biomarker comparison.
\newblock \emph{BMC Medical Research Methodology}, 17, 2017.

\bibitem[Atiquzzaman et~al.(2020)Atiquzzaman, Karim, Kopec, Wong, Vera, and
  Anis]{Atiquzzaman2020}
M.~Atiquzzaman, M.~E. Karim, J.~Kopec, H.~Wong, M.~A. Vera, and A.~H. Anis.
\newblock Using external data to incorporate unmeasured confounders: A plasmode
  simulation study comparing alternative approaches to impute body max index in
  a study of the relationship between osteoarthritis and cardiovascular
  disease.
\newblock \emph{Journal of Statistical Research}, 54(2):\penalty0 131--145,
  2020.

\bibitem[Shmueli(2010)]{Shmueli2010}
G.~Shmueli.
\newblock To explain or to predict?
\newblock \emph{Statistical Science}, 25:\penalty0 289–310, 2010.

\bibitem[Mehta et~al.(2004)Mehta, Tanik, and Allison]{Mehta2004}
T.~Mehta, M.~Tanik, and D.~B. Allison.
\newblock Towards sound epistemological foundations of statistical methods for
  high-dimensional biology.
\newblock \emph{Nature genetics}, 36:\penalty0 943--947, 2004.

\bibitem[Boulesteix et~al.(2020)Boulesteix, Groenwold, Abrahamowicz, Binder,
  Briel, Hornung, Morris, Rahnenführer, and Sauerbrei]{Boulesteix2020}
A.-L. Boulesteix, R.~H. Groenwold, M.~Abrahamowicz, H.~Binder, M.~Briel,
  R.~Hornung, T.~P. Morris, J.~Rahnenführer, and W.~Sauerbrei.
\newblock Introduction to statistical simulations in health research.
\newblock \emph{BMJ Open}, 2020.

\bibitem[Mehta et~al.(2006)Mehta, Zakharkin, Gadbury, and Allison]{Mehta2006}
T.~Mehta, S.~O. Zakharkin, G.~L. Gadbury, and D.~B. Allison.
\newblock Epistemological issues in omics and high-dimensional biology: give
  the people what they want.
\newblock \emph{Physiol Genomics}, 28:\penalty0 24--32, 2006.

\bibitem[Gadbury et~al.(2008)Gadbury, Xiang, Yang, Barnes, Page, and
  Allison]{Gadbury2008}
G.~L. Gadbury, Q.~Xiang, L.~Yang, S.~Barnes, G.~P. Page, and D.~B. Allison.
\newblock Evaluating statistical methods using plasmode data sets in the age of
  massive public databses: an illustration using false discovery rates.
\newblock \emph{{PL}o{S} Genetics}, 6, 2008.

\bibitem[Vaughan et~al.(2009)Vaughan, Divers, Padilla, Redden, Tiwari, Pomp,
  and Allison]{Vaughan2009}
L.~K. Vaughan, J.~Divers, M.~A. Padilla, D.~T. Redden, H.~K. Tiwari, D.~Pomp,
  and D.~B. Allison.
\newblock The use of plasmodes as a supplement to simulations: A simple example
  evaluating individual admixture estimation methodologies.
\newblock \emph{Computational Statistics and Data Analysis}, 53:\penalty0
  1755--1766, 2009.

\bibitem[Headrick and Sawilowsky(1999)]{Headrick1999}
T.~C. Headrick and S.~S. Sawilowsky.
\newblock Simulating correlated multivariate nonnormal distributions: extending
  the fleschman power method.
\newblock \emph{Psychometrika}, 64\penalty0 ((1)):\penalty0 25--35, 1999.

\bibitem[Azuero et~al.(2012)Azuero, Redden, Tiwari, Asmelash, and
  Piyathilake]{Azuero2012}
A.~Azuero, D.~T. Redden, H.~K. Tiwari, S.~G. Asmelash, and C.~J. Piyathilake.
\newblock A simple distribution-free algorithm for generating simulated
  high-dimensional correlated data with an autoregressive structure.
\newblock \emph{Commun Stat Simul Comput}, 41(1):\penalty0 89--98, 2012.

\bibitem[Fan et~al.(2014)Fan, Han, and Liu]{Fan2014}
J.~Fan, F.~Han, and H.~Liu.
\newblock Challenges of big data analysis.
\newblock \emph{National Science Review}, 1\penalty0 ((2)):\penalty0 293--314,
  2014.

\bibitem[Fan and Li(2006)]{Fan2006}
J.~Fan and R.~Li.
\newblock Statistical challenges with high dimensionality: feature selection in
  knowledge discovery.
\newblock \emph{Proceedings of the International Congress of Mathematicians,
  Madrid, Spain}, pages 595--622, 2006.

\bibitem[Johnstone and Titterington(2009)]{Johnstone2009}
I.~M. Johnstone and D.~M. Titterington.
\newblock Statistical challenges of high-dimensional data.
\newblock \emph{Phil. Trans. R. Soc. A}, 367:\penalty0 4237–4253, 2009.

\bibitem[Cattell and Jaspers(1967)]{Cattell1967}
R.~Cattell and J.~Jaspers.
\newblock A general plasmode (no. 30-10-5-2) for factor analytic exercises and
  research.
\newblock 67-3, 1967.

\bibitem[Sokal et~al.(1980)Sokal, Rohlf, Zang, and Osness]{Sokal1980}
R.~R. Sokal, F.~J. Rohlf, E.~Zang, and W.~Osness.
\newblock Reification in factor analysis: a plasmode based on human
  physiology-of-exercise variables.
\newblock \emph{Multivariate Behavioral Research}, 4, 1980.

\bibitem[Irizarry et~al.(2003)Irizarry, Hobbs, Collin, Beazer-Barclay,
  Antonellis, Scherf, and Speed]{Irizarry2003}
R.~A. Irizarry, B.~Hobbs, F.~Collin, Y.~D. Beazer-Barclay, K.~J. Antonellis,
  U.~Scherf, and T.~P. Speed.
\newblock Exploration, normalization, and summaries of high density
  oligonucleotide array probe level data.
\newblock \emph{Biostatistics}, 4:\penalty0 249--264, 2003.

\bibitem[Tibshirani(2006)]{Tibshirani2006}
R.~Tibshirani.
\newblock A simple method for assessing sample sizes in microarray experiments.
\newblock \emph{BMC Bioinformatics}, 7, 2006.

\bibitem[Franklin et~al.(2015)Franklin, Eddings, Glynn, and
  Schneeweiss]{Franklin2015}
J.~M. Franklin, W.~Eddings, R.~J. Glynn, and S.~Schneeweiss.
\newblock Regularized regression versus the high-dimensional propensity score
  for confounding adjustment in secondary database analyses.
\newblock \emph{American Journal of Epidemiology}, 182\penalty0 ((7)):\penalty0
  651--659, 2015.

\bibitem[Franklin et~al.(2017)Franklin, Eddings, Austin, Stuart, and
  Schneeweiss]{Franklin2017}
J.~M. Franklin, W.~Eddings, P.~C. Austin, E.~A. Stuart, and S.~Schneeweiss.
\newblock Comparing the performance of propensity score methods in healthcare
  database studies with rare outcomes.
\newblock \emph{Statistics in Medicine}, 36:\penalty0 1946--1963, 2017.

\bibitem[Karim et~al.(2018)Karim, Pang, and Platt]{Karim2021}
M.~E. Karim, M.~Pang, and R.~W. Platt.
\newblock Can we train machine learning methods to outperform the
  high-dimensional propensity score algorithm.
\newblock \emph{Epidemiology}, 2018.

\bibitem[Desai et~al.(2019)Desai, Wyss, Abdia, Toh, Johnson, Lee, Karami,
  Major, Nguyen, Wang, Franklin, and Gagne]{Desai2019}
R.~J. Desai, R.~Wyss, Y.~Abdia, S.~Toh, M.~Johnson, H.~Lee, S.~Karami, J.~M.
  Major, M.~Nguyen, S.~V. Wang, J.~M. Franklin, and J.~J. Gagne.
\newblock Evaluating the use of bootstrapping in cohort studies conducted with
  1:1 propensity score matching - a plasmode simulation study.
\newblock \emph{Pharmacoepidemiol Drug Saf}, 28:\penalty0 879--886, 2019.

\bibitem[Ripollone et~al.(2019)Ripollone, Huybrechts, Rothman, Ferguson, and
  Franklin]{Ripollone2019}
J.~E. Ripollone, K.~F. Huybrechts, K.~J. Rothman, R.~E. Ferguson, and J.~M.
  Franklin.
\newblock Evaluating the utility of coarsened exact matching for
  pharmacoepidemiology using real and simulated claims data.
\newblock \emph{Practice of Epidemiology}, 189\penalty0 ((6)):\penalty0
  613--622, 2019.

\bibitem[Liu et~al.(2019)Liu, Chrysanthopoulou, Chang, Hunnicutt, and
  Lapane]{Liu2019}
S.-H. Liu, S.~A. Chrysanthopoulou, Q.~Chang, J.~N. Hunnicutt, and K.~L. Lapane.
\newblock Missing data in marginal structural models: A plasmode simulation
  study comparing multiple imputation and inverse probability weighting.
\newblock \emph{Medical Care}, 57\penalty0 ((3)):\penalty0 237--243, 2019.

\bibitem[Conover et~al.(2021)Conover, Rothman, St{\"u}rmer, Ellis, Poole, and
  Funk]{Conover2021}
M.~M. Conover, K.~J. Rothman, T.~St{\"u}rmer, A.~R. Ellis, C.~Poole, and M.~J.
  Funk.
\newblock Propensity score trimming mitigates bias due to covariate measurement
  error in inverse probability of treatment weighted analyses: A plasmode
  simulation.
\newblock \emph{Statistics in Medicine}, 40:\penalty0 2101--2112, 2021.

\bibitem[Wyss et~al.(2021)Wyss, Schneeweiss, van~der Laan, Lendle, Ju, and
  Franklin]{Wyss2021}
R.~Wyss, S.~Schneeweiss, M.~van~der Laan, S.~D. Lendle, C.~Ju, and J.~M.
  Franklin.
\newblock Using super learner prediction modeling to improve high-dimensional
  propensity score estimation.
\newblock \emph{Epidemiology}, 2021.

\bibitem[Hafermann et~al.(2022)Hafermann, Klein, Rauch, Kammer, and
  Heinze]{Hafermann2022}
L.~Hafermann, N.~Klein, G.~Rauch, M.~Kammer, and G.~Heinze.
\newblock Using background knowledge from preceding studies for building a
  random forest prediction model: A plasmode simulation study.
\newblock \emph{Entropy}, 24, 2022.

\bibitem[Rodriguez et~al.(2022)Rodriguez, Veenstra, Heagerty, Goss, Ramos, and
  Bansal]{Rodriguez2022}
P.~J. Rodriguez, D.~L. Veenstra, P.~J. Heagerty, C.~H. Goss, K.~J. Ramos, and
  A.~Bansal.
\newblock A framework for using real-world data and health outcomes modeling to
  evaluate machine learning based risk prediction models.
\newblock \emph{Value Health}, 25\penalty0 (3):\penalty0 350–358, 2022.

\bibitem[Elobeid et~al.(2009)Elobeid, Padilla, McVie, Thomas, Brock, Musser,
  Lu, Coffey, Desmond, St-Onge, Gadde, Heymsfield, and Allison]{Elobeid2009}
M.~A. Elobeid, M.~A. Padilla, T.~McVie, O.~Thomas, D.~W. Brock, B.~Musser,
  K.~Lu, C.~S. Coffey, R.~A. Desmond, M.-P. St-Onge, K.~M. Gadde, S.~B.
  Heymsfield, and D.~B. Allison.
\newblock Missing data in randomized clinical trials for weight loss: Scope of
  the problem, state of the field, and performance of statistical methods.
\newblock \emph{PLoS One}, 4\penalty0 ((8)), 2009.

\bibitem[Reeb et~al.(2015)Reeb, Bramardi, and Steibel]{Reeb2015}
P.~D. Reeb, S.~J. Bramardi, and J.~P. Steibel.
\newblock Assessing dissimilarity measures for sample based hierarchical
  clustering of rna sequencing data using plasmode datasets.
\newblock \emph{PLOS Genetics}, 2015.

\bibitem[Ejima et~al.(2020)Ejima, Brown, Smith, Beyaztas, and
  Allison]{Ejima2020}
K.~Ejima, A.~W. Brown, D.~L. Smith, U.~Beyaztas, and D.~B. Allison.
\newblock Murine genetic models of obesity: type i error rates and the power of
  commonly used analyses as assessed by plasmode-based simulation.
\newblock \emph{International Journal of Obesity}, 44:\penalty0 1440--1449,
  2020.

\bibitem[Alfaras et~al.(2021)Alfaras, Ejima, Teixeira, Germanio, Mitchell,
  Hamilton, Ferrucci, Price, Allison, Bernier, and de~Cabo]{Alfaras2021}
I.~Alfaras, K.~Ejima, C.~V.~L. Teixeira, C.~D. Germanio, S.~J. Mitchell,
  S.~Hamilton, L.~Ferrucci, N.~L. Price, D.~B. Allison, M.~Bernier, and
  R.~de~Cabo.
\newblock Empirical versus theoretical power and type i error (false-positive)
  rates estimated from real murine aging research data.
\newblock \emph{Cell Reports}, 36(7), 2021.

\bibitem[Bickel et~al.(1997)Bickel, G{\"o}tze, and van Zweet]{Bickel1997}
B.~J. Bickel, F.~G{\"o}tze, and W.~R. van Zweet.
\newblock Resampling fewer than n observations: gains, losses, and remedies for
  losses.
\newblock \emph{Statistica Sinica}, 7:\penalty0 1--31, 1997.

\bibitem[Bickel and Sakov(2008)]{Bickel2008}
B.~J. Bickel and A.~Sakov.
\newblock On the choice of m in the m out of n bootstrap and confidence bounds
  for extrema.
\newblock \emph{Statistica Sinica}, 18:\penalty0 967--985, 2008.

\bibitem[Andrews and Buchinsky(2000)]{Andrews2000}
D.~W.~K. Andrews and M.~Buchinsky.
\newblock A three-step method for choosing the number of bootstrap repetitions.
\newblock \emph{Econometrica}, 68:\penalty0 23–51, 2000.

\bibitem[Davidson and MacKinnon(2000)]{Davidson2000}
R.~Davidson and J.~G. MacKinnon.
\newblock Bootstrap tests: How many bootstraps?
\newblock \emph{Econometric Reviews}, 19:\penalty0 55–68, 2000.

\bibitem[Politis et~al.(1999)Politis, Romano, and Wolf]{Politis1999}
D.~Politis, J.~Romano, and M.~Wolf.
\newblock \emph{Subsampling}.
\newblock Springer, 1999.

\bibitem[Bickel and Ren(2001)]{Bickel2001}
B.~J. Bickel and J.-J. Ren.
\newblock The bootstrap in hypothesis testing.
\newblock \emph{Lecture Notes-Monograph Series, State of the Art in Probability
  and Statistics}, 36:\penalty0 91--112, 2001.

\bibitem[Andrews and Guggenberger(2010)]{Andrews2010}
D.~W.~K. Andrews and P.~Guggenberger.
\newblock Asymptotic size and a problem with subsampling and with the m out of
  n bootstrap.
\newblock \emph{Econometric Theory}, 26:\penalty0 426–468, 2010.

\bibitem[Gerard(2020)]{Gerard2020}
D.~Gerard.
\newblock Data-based rna-seq simulations by binomial thinning.
\newblock \emph{BMC Boinformatics}, 21:\penalty0 206--220, 2020.

\bibitem[Ju et~al.(2019)Ju, Wyss, Franklin, Schneeweiss, H{\"a}ggstr{\"o}m, and
  van~der Laan]{Ju2019}
C.~Ju, R.~Wyss, J.~M. Franklin, S.~Schneeweiss, J.~H{\"a}ggstr{\"o}m, and M.~J.
  van~der Laan.
\newblock Collaborative-controlled {LASSO} for constructing propensity
  score-based estimators in high-dimensional data.
\newblock \emph{Statistical Methods in Medical Research}, 28:\penalty0
  1044--1063, 2019.

\bibitem[Beran and Srivastava(1985)]{Beran1985}
R.~Beran and M.~S. Srivastava.
\newblock Bootstrap tests and confidence regions for functions of a covariance
  matrix.
\newblock \emph{The Annals of Statistics}, 13:\penalty0 95–115, 1985.

\bibitem[Bickel and Freedman()]{Bickel1983}
P.~J. Bickel and D.~A. Freedman.
\newblock Bootstrapping regression models with many parameters.
\newblock In \emph{A Festschrift for Erich L. Lehmann in Honor of his
  Sixty-fith Birthday}.

\bibitem[Mammen(1993)]{Mammen1993}
E.~Mammen.
\newblock Bootstrap and wild bootstrap for high dimensional linear models.
\newblock \emph{The Annals of Statistics}, 21:\penalty0 255–285, 1993.

\bibitem[Karoui and Purdom(2018)]{ElKaroui2018}
N.~E. Karoui and E.~Purdom.
\newblock Can we trust the bootstrap in high-dimensions? the case of linear
  models.
\newblock \emph{Journal of Machine Learning Research}, 19:\penalty0 1--66,
  2018.

\bibitem[Hoerl and Kennard(1970)]{Hoerl1970}
A.~E. Hoerl and R.~W. Kennard.
\newblock Ridge regression: Biased estimation for nonorthogonal problems.
\newblock \emph{Technometrics}, 12:\penalty0 55–67, 1970.

\bibitem[Searle et~al.(1992)Searle, Casella, and McCulloch]{Searle1992}
S.~R. Searle, G.~Casella, and C.~E. McCulloch.
\newblock \emph{Variance Components}.
\newblock Wiley Interscience, New Jersey, 1992.

\bibitem[Covarrubias-Pazaran(2017)]{Pazar2017}
G.~Covarrubias-Pazaran.
\newblock Genome assisted prediction of quantitative traits using the {R}
  package sommer.
\newblock \emph{PLoS ONE}, 11:\penalty0 1--15, 2017.

\bibitem[Weinstein et~al.(2013)Weinstein, Collisson, and et.
  al.]{Weinstein2013}
J.~Weinstein, E.~Collisson, and et. al.
\newblock The cancer genome atlas pan-cancer analysis project.
\newblock \emph{Nature Genetics}, 45:\penalty0 1113–1120, 2013.
\newblock URL \url{https://doi.org/10.1038/ng.2764}.

\bibitem[Colaprico et~al.(2015)Colaprico, Silva, Olsen, Garofano, Cava,
  Garolini, Sabedot, Malta, Pagnotta, Castiglioni, Ceccarelli, Bontempi, and
  Noushmehr]{Colaprico2015}
A.~Colaprico, T.~C. Silva, C.~Olsen, L.~Garofano, C.~Cava, D.~Garolini,
  T.~Sabedot, T.~M. Malta, S.~M. Pagnotta, I.~Castiglioni, M.~Ceccarelli,
  G.~Bontempi, and H.~Noushmehr.
\newblock Tcgabiolinks: An r/bioconductor package for integrative analysis of
  tcga data.
\newblock \emph{Nucleic Acids Research}, 2015.
\newblock \doi{10.1093/nar/gkv1507}.
\newblock URL \url{http://doi.org/10.1093/nar/gkv1507}.

\bibitem[{Silva} et~al.(2016){Silva}, C, {Colaprico}, {Antonio}, {Olsen},
  {Catharina}, {D'Angelo}, {Fulvio}, {Bontempi}, {Gianluca}, {Ceccarelli},
  {Michele}, {Noushmehr}, and {Houtan}]{Silva2016}
{Silva}, T.~C, {Colaprico}, {Antonio}, {Olsen}, {Catharina}, {D'Angelo},
  {Fulvio}, {Bontempi}, {Gianluca}, {Ceccarelli}, {Michele}, {Noushmehr}, and
  {Houtan}.
\newblock Tcga workflow: Analyze cancer genomics and epigenomics data using
  bioconductor packages.
\newblock \emph{F1000Research}, 5, 2016.

\bibitem[{Mounir} et~al.(2019){Mounir}, {Mohamed}, {Lucchetta}, {Marta},
  {Silva}, C, {Olsen}, {Catharina}, {Bontempi}, {Gianluca}, {Chen}, {Xi},
  {Noushmehr}, {Houtan}, {Colaprico}, {Antonio}, {Papaleo}, and
  {Elena}]{Mounir2019}
{Mounir}, {Mohamed}, {Lucchetta}, {Marta}, {Silva}, T.~C, {Olsen}, {Catharina},
  {Bontempi}, {Gianluca}, {Chen}, {Xi}, {Noushmehr}, {Houtan}, {Colaprico},
  {Antonio}, {Papaleo}, and {Elena}.
\newblock New functionalities in the tcgabiolinks package for the study and
  integration of cancer data from gdc and gtex.
\newblock \emph{PLoS computational biology}, 15\penalty0 (3):\penalty0
  e1006701, 2019.

\bibitem[Ritchie et~al.(2015)Ritchie, Phipson, Wu, Hu, Law, Shi, and
  Smyth]{Ritchie2015}
M.~E. Ritchie, B.~Phipson, D.~Wu, Y.~Hu, C.~W. Law, W.~Shi, and G.~K. Smyth.
\newblock {limma} powers differential expression analyses for {RNA}-sequencing
  and microarray studies.
\newblock \emph{Nucleic Acids Research}, 43\penalty0 (7):\penalty0 e47, 2015.
\newblock \doi{10.1093/nar/gkv007}.

\bibitem[Ledoit and Wolf(2004)]{Ledoit2004}
O.~Ledoit and M.~Wolf.
\newblock A well-conditioned estimator for large-dimensional covariance
  matrices.
\newblock \emph{Journal of Multivariate Analysis}, 88\penalty0 (2):\penalty0
  365--411, 2004.
\newblock
  \doi{https://www.sciencedirect.com/science/article/pii/S0047259X03000964}.

\bibitem[Tibshirani(1996)]{Tibshirani1996}
R.~Tibshirani.
\newblock Regression shrinkage and selection via the lasso.
\newblock \emph{Journal of the Royal Statistical Society B}, 58:\penalty0
  267–288, 1996.

\end{thebibliography}

\clearpage

\end{document}